\RequirePackage{lineno}

\documentclass[aps,preprint,showpacs,superscriptaddress,groupedaddress,amsmath]{revtex4}

\usepackage{amsmath,amsfonts,graphicx,bm,amssymb,array}
\usepackage[usenames]{color}
\usepackage{array}
\usepackage{float}
\usepackage{sidecap}
\usepackage{graphicx}
\usepackage{epstopdf}
\usepackage{color}
\usepackage{subfigure}
\usepackage{longtable}

\def\etmiss{E\!\!\!\!\slash_{T}}

\begin{document}

\title{Single top quark production in the $t$-channel at 14~TeV and 33~TeV}

\author{Brad Schoenrock} \affiliation{Department of Physics and Astronomy, Michigan State University, East Lansing MI 48824, USA}
\author{Elizabeth Drueke} \affiliation{Department of Physics and Astronomy, Michigan State University, East Lansing MI 48824, USA}
\author{Barbara Alvarez Gonzalez} \affiliation{Department of Physics and Astronomy, Michigan State University, East Lansing MI 48824, USA}
\author{Reinhard Schwienhorst} \affiliation{Department of Physics and Astronomy, Michigan State University, East Lansing MI 48824, USA}

\date{\today}

\begin{abstract}
We study $t$-channel single top quark production at future LHC runs at 14~TeV with
300~fb$^{-1}$ and 3000~fb$^{-1}$ as well as at a
future 33~TeV proton-proton collider in the context of the Snowmass 2013 study.
The single top final state has a lepton and neutrino from the top
quark decay plus two jets, one of which is required to be
$b$-tagged. We show that it is possible to isolate large samples of single top events and
that the cross-section can be measured with a precision of 5\% or better.
\end{abstract}

\maketitle

\modulolinenumbers[1]
\linenumbers


\section{Introduction}
\label{sec:intro}
~
The Large Hadron Collider (LHC) is the highest-energy particle accelerator ever built,
probing physics at the TeV scale. The Higgs boson discovery~\cite{Aad:2012tfa,Chatrchyan:2012ufa}
was the first, but more discoveries are likely at the LHC which covers the energy range
where new physics is anticipated. 

\begin{figure}[!h!tpb]
  \includegraphics[width=0.35\textwidth]{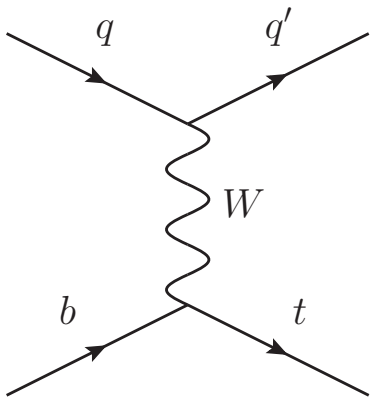}
  \caption{Leading order Feynman diagram for single top quark production in the $t$-channel.}
  \label{fig:tchan_feynman}
\end{figure}

Single top quarks are produced in the Standard Model (SM) through three different mechanisms: 
the $t$-channel exchange of a $W$~boson, the corresponding 
Feynman diagram is shown in Fig.~\ref{fig:tchan_feynman}, 
the associated production of an on-shell $W$~boson and a top-quark,
and the $s$-channel production and decay of a virtual $W$~boson.

The single top quark discovery has been reported by the Tevatron experiments~\cite{Group:2009qk,Aaltonen:2009jj,Abazov:2009ii}, 
based on a combination of $t$-channel and $s$-channel processes in 2009. 
The $t$-channel process  has been observed at the Tevatron~\cite{CDF-SGTOP-7.5,Abazov:2011rz} and at the 7 TeV LHC 
by ATLAS~\cite{Aad:2012ux} and CMS~\cite{CMS:2012ep, CMS:2013tch}. 
The current precision of the $t$-channel cross-section measurements at 8~TeV performed by
ATLAS~\cite{ATLAS8TeVtchan} and CMS~\cite{CMS8TeVtchan} is about 15$\%$. Work is in progress
to update these results with the full 2012 8~TeV dataset.

In this paper we explore the precision with which it will be possible to measure
the single top quark production cross-section in the $t$-channel for 
three different scenarios: 300~fb$^{-1}$ of 14~TeV data
with an average pileup of 50~events, also known as the Phase-1 running of the LHC upgrade;
3000~fb$^{-1}$ of 14~TeV data with an average pileup of 140~events, also known as Phase-2
running or the high-luminosity LHC (HL-LHC); and 3000~fb$^{-1}$ of 33~TeV data with 
an average pileup of 140~events, also known as the high-energy LHC (HE-LHC).
For each scenario we use the appropriate Snowmass detector model.
The analysis is performed
using a simple cut-based approach to extract the $t$-channel signal. It is tuned to get at
least a signal to background ratio of five in order to obtain a clean sample of $t$-channel
single-top events. 

This paper is organized as follows:
Section~\ref{sec:models} describes the Monte Carlo samples that are used for this analysis.
Section~\ref{sec:analysis} explains the event selection and the analysis strategy. 
Section~\ref{sec:selection} shows the final event selection and
Section~\ref{sec:conclusions} gives our conclusions.

%
\section{Signal and Background modeling}
\label{sec:models}
The signal $t$-channel events are generated with Madgraph5~\cite{Alwall:2011uj,MadAnalysis},
with Pythia8~\cite{pythia,pythiapgs} for parton showering and Delphes~\cite{delphes3} for
modeling of the Snowmass LHC detector~\cite{Snowmasstwiki}. 
We consider inclusive $t$-channel events $\overline{b}~t$ and $b~\overline{t}$ and
no attempt is made to separate top from antitop.

Background samples for W/$\gamma$/Z+Jets, diboson, $t\bar{t}$, and $Wt$ single top are used
from the official Snowmass webpage~\cite{Snowmasstwiki}. These samples are generated in bins of
$H_T$, each with its own cross-section.
Normalizations are made according to the cross-sections provided by the Snowmass site for the
background samples and according to the leading order cross-sections provided by MadGraph for
the signal samples. The total cross-section for the signal and backgrounds at 14~TeV and
33~TeV are listed in Table~\ref{tab:xsecs} and~\ref{tab:xsecs33TeV}, respectively.

\begin{table}[!h!tbp]
    \begin{tabular}{|l|c|}
      \hline
      Sample     &     Cross-Section [pb] \\
      \hline
      t-channel  & 30.2\\
      \hline
      $W$+jets   & 238400\\
      \hline
      $t\bar{t}$ & 578.5\\
      \hline
      Diboson    & 289.8\\
      \hline
      $Wt$       & 72.1\\
      \hline
      
    \end{tabular}
    \caption{cross-sections for each background sample at 14~TeV.} 
    \label{tab:xsecs}
  \end{table}
\begin{table}[h!]
    \begin{tabular}{|l|c|}
      \hline
      Sample     &     Cross-Section [pb] \\
      \hline
      t-channel  & 493.5 \\
      \hline
      $W$+jets   & 674860.6\\
      \hline
      $t\bar{t}$ & 4014.8\\
      \hline
      Diboson    & 894.0\\
      \hline
      $Wt$       & 492.6\\
      \hline     
     \end{tabular}
    \caption{cross-sections for each background sample at 33~TeV.} 
    \label{tab:xsecs33TeV}
  
\end{table}
\section{Analysis}
\label{sec:analysis}

For this analysis a cut-and-count method is used. To account for differences in 
generation, select the relevant event signature, and to provide object definitions 
the following basic event selection is applied, on which subsequent steps are based: 
~
\begin{eqnarray}
\textrm{One lepton (electron or muon) with }& \qquad p_{T}^{\ell}&\geq 40\,{\rm GeV}, \qquad
\left|\eta_{\ell}\right|\leq 2.5, \nonumber \\
\textrm{Two jets with }& \qquad p_T^j&\geq 70\,{\rm GeV}, \qquad 
\left|\eta_{j}\right|\leq 4.5, \nonumber \\
\textrm{One b-jet with }& \qquad p_T^j&\geq 70\,{\rm GeV}, \qquad 
\left|\eta_{j}\right|\leq 2.5, \nonumber \\
\textrm{Missing energy }& \qquad  ~\etmiss &\geq 30~{\rm GeV}  \nonumber \\
\label{eq:basiccut}
\end{eqnarray}
where $p_{T}^{\ell}$ and $\eta_{\ell}$ correspond to the transverse momentum and pseudorapidity of the lepton, respectively, 
and $p_T^j$ and $\eta_{j}$ are the transverse momentum and pseudorapidity of each jet. For 33 TeV $p_{T}^{\ell}\geq$ 50 GeV 
and $p_T^j \geq$ 75 GeV are used.

The first line in tables~\ref{tab:yields50}, ~\ref{tab:yields140} and~\ref{tab:yields33} 
give the expected event yields for the signal and backgrounds with basic event selection
for the different scenarios.
At this stage the ratio of the signal contribution over the total background is about 0.13, 
with $t\bar{t}$ being the dominant background contribution.

\begin{figure}[!h!tbp]
  \centering
  \subfigure[]{
    \includegraphics[width=0.40\textwidth]{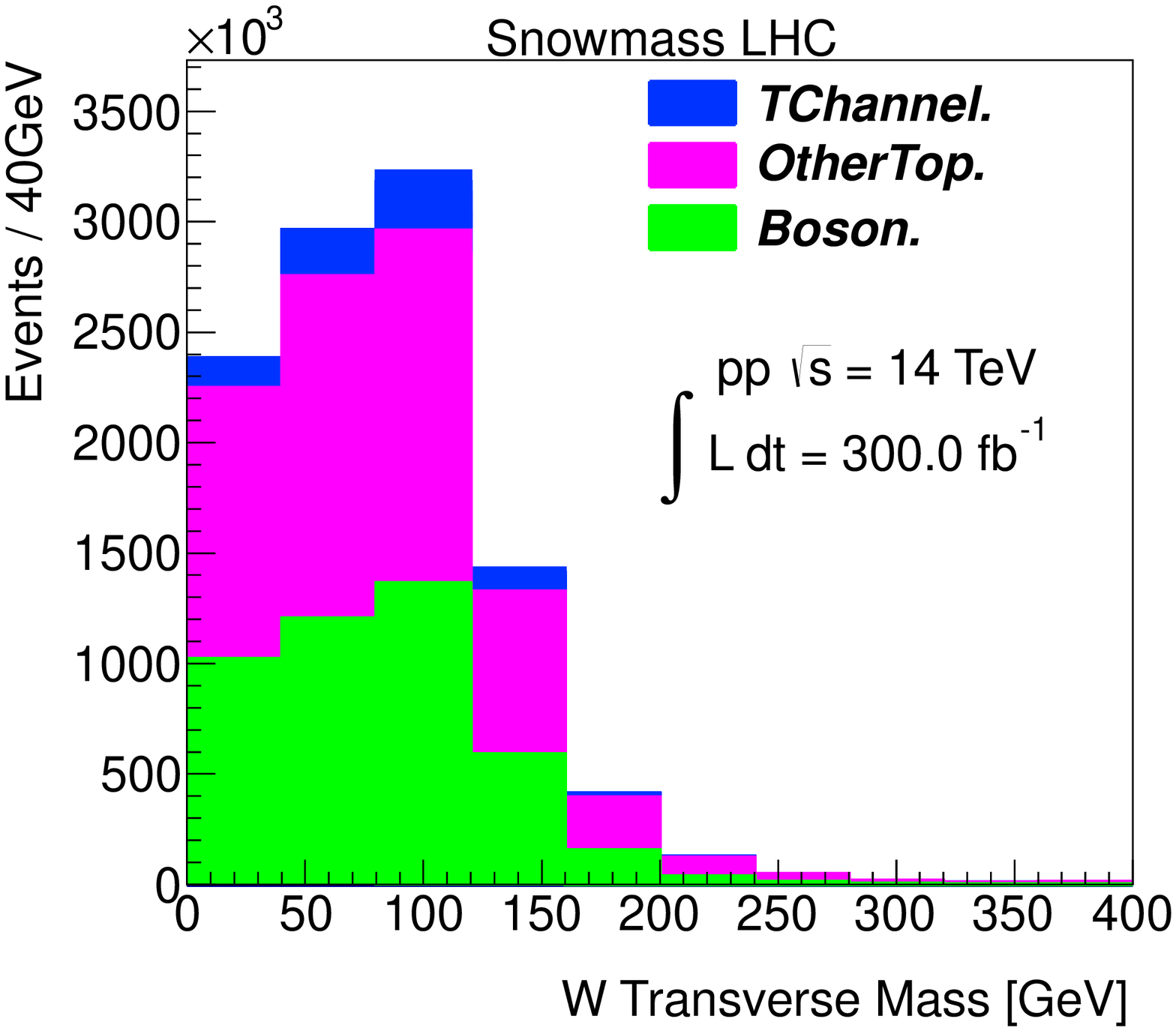}
    \label{fig:a}
  }
  \hspace*{0.0\textwidth}
  \subfigure[]{
    \includegraphics[width=0.40\textwidth]{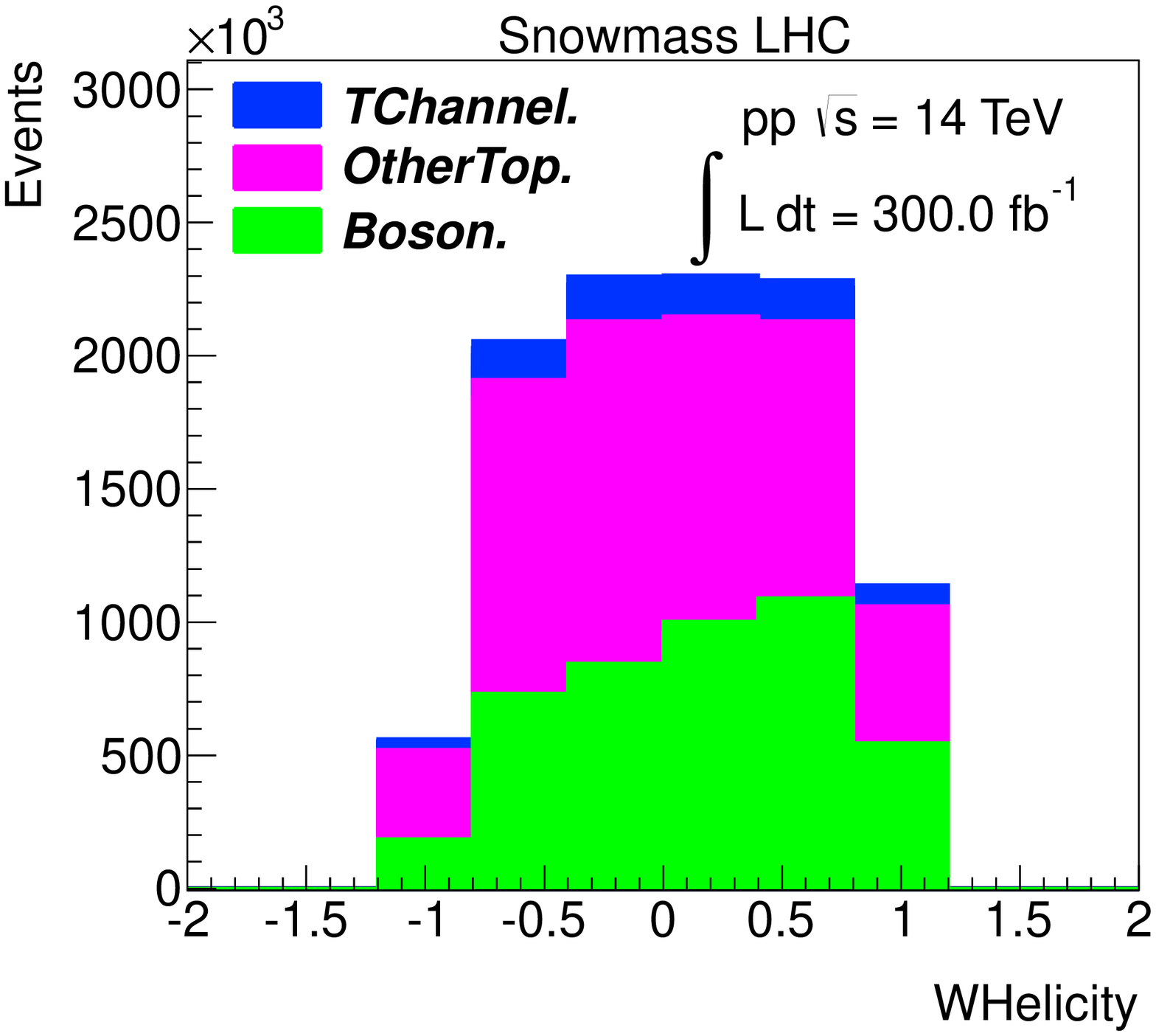}
    \label{fig:b}
  }
  \subfigure[]{
    \includegraphics[width=0.40\textwidth]{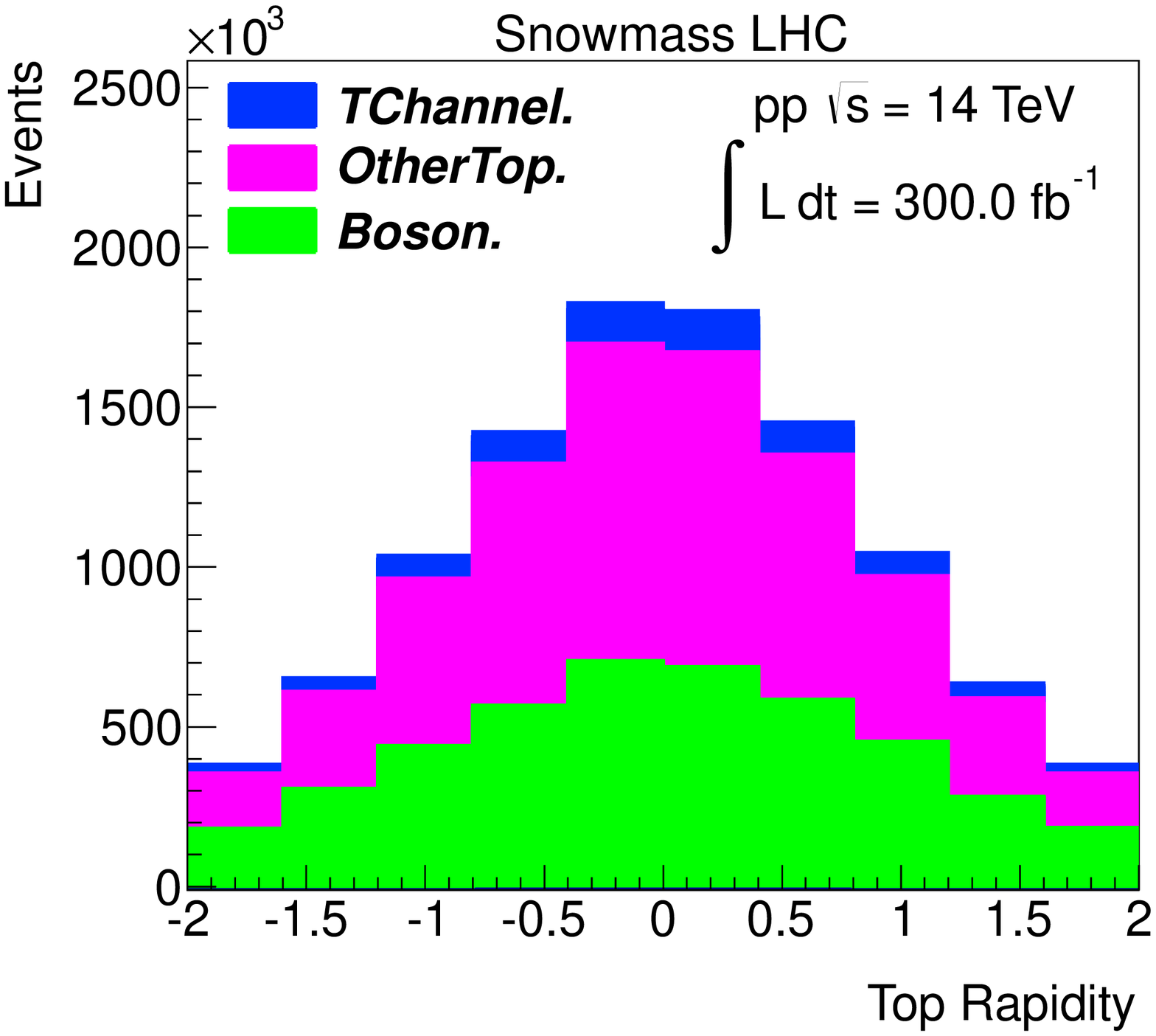}
    \label{fig:c}
  } 
  \hspace*{0.0\textwidth}
  \subfigure[]{
    \includegraphics[width=0.40\textwidth]{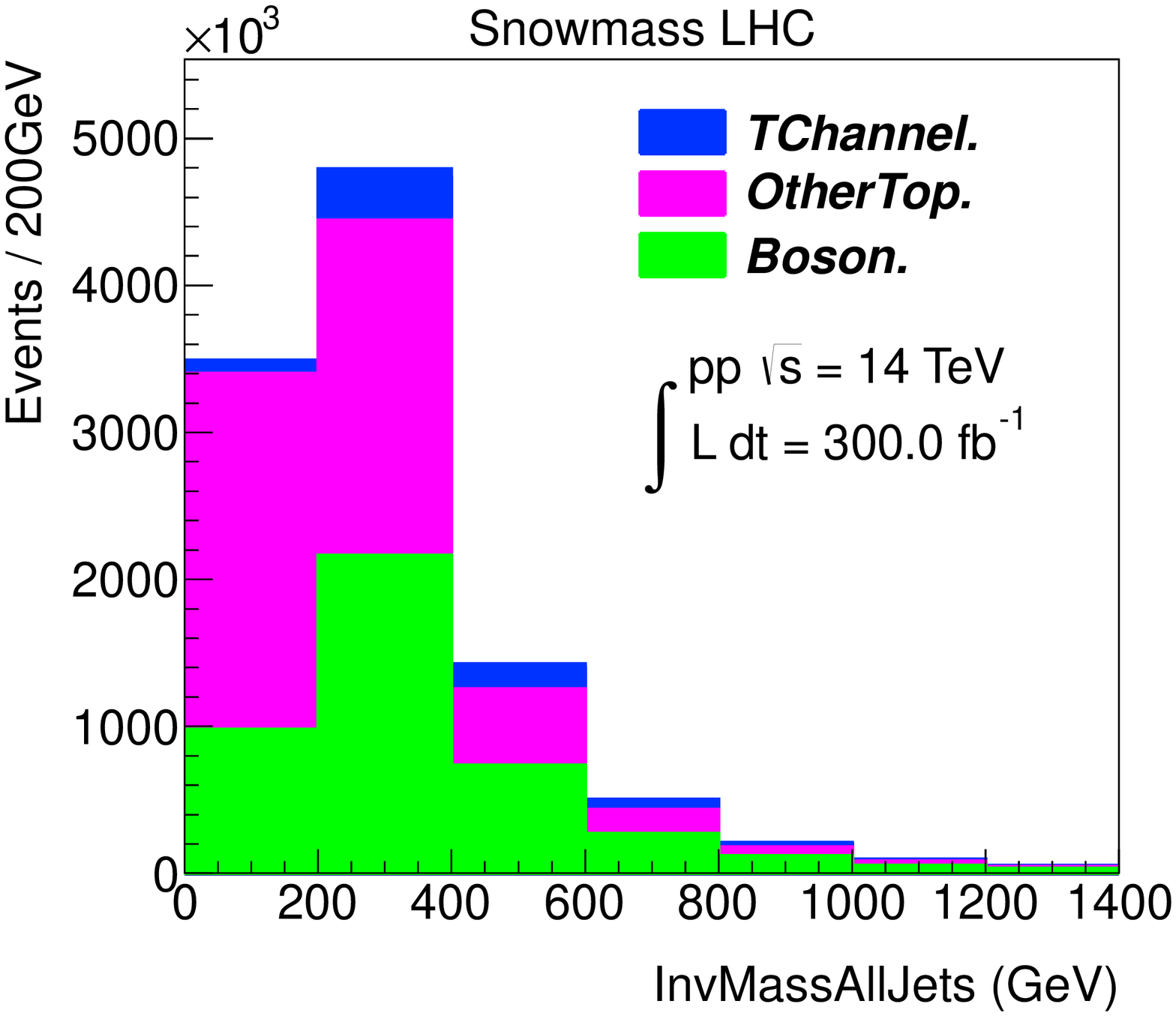}
    \label{fig:d}
  } 
  \caption{Kinematic distributions for 300~fb$^{-1}$ at 14~TeV: (a) transverse mass of the W boson, (b) helicity of the W boson, (c) rapidity of the top quark  and (d) invariant mass of all the jets in the event.}
  \label{fig:prekinematics}
\end{figure}

\begin{figure}
  \centering
  \subfigure[]{
    \includegraphics[width=0.40\textwidth]{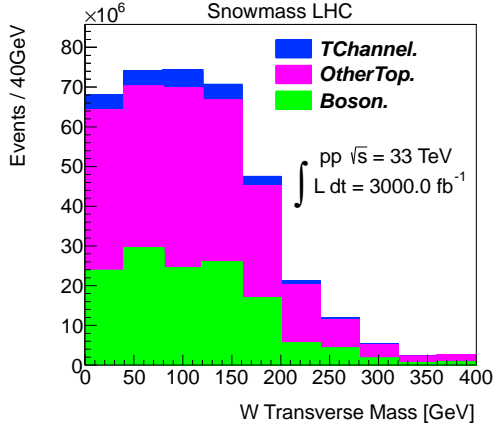}
    \label{fig:a}
  }
  \hspace*{0.0\textwidth}
  \subfigure[]{
    \includegraphics[width=0.40\textwidth]{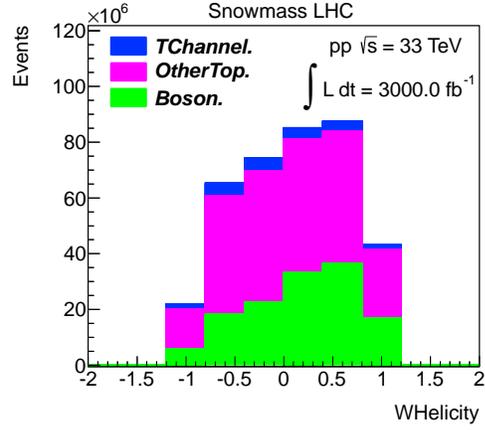}
    \label{fig:b}
  }
  \subfigure[]{
    \includegraphics[width=0.40\textwidth]{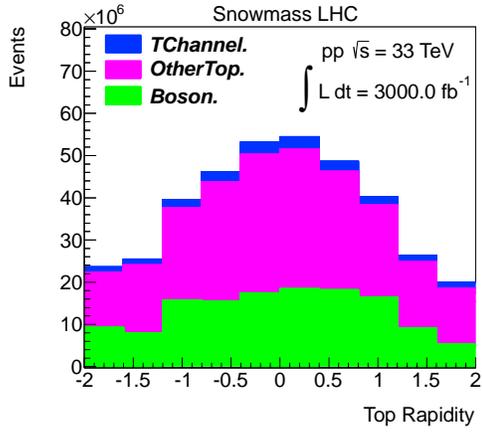}
    \label{fig:c}
  } 
  \hspace*{0.0\textwidth}
  \subfigure[]{
    \includegraphics[width=0.40\textwidth]{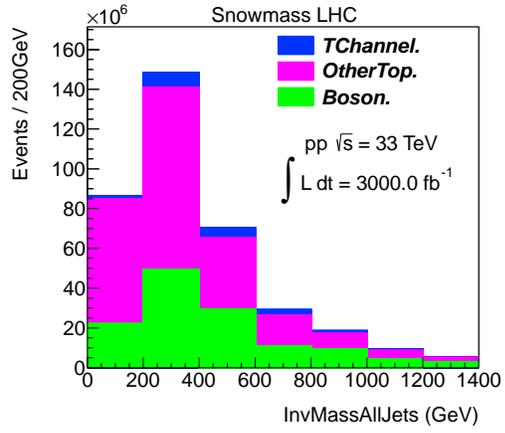}
    \label{fig:d}
  } 
  \caption{Kinematic distributions for 3000~fb$^{-1}$ at 33~TeV: (a) transverse mass of the W boson, (b) helicity of the W boson, (c) rapidity of the top quark  and (d) invariant mass of all the jets in the event.}
  \label{fig:prekinematics33}
\end{figure}

Figure~\ref{fig:prekinematics} shows some kinematic distributions after the preselection cuts are applied. 
The contributions are broken into the $t$-channel single top signal events, other top events that correspond with 
$Wt$ and $t\bar{t}$, and other backgrounds that come from $W$+jets and diboson events (labeled as Boson on the plots). 
The expected events are normalized using the 14~TeV reference cross-sections and the integrated 
luminosity of 300~fb$^{-1}$. The distributions for 33~TeV are shown in Figure~\ref{fig:prekinematics33}. 
~

\section{Event selection}
\label{sec:selection}

To improve signal to background ratio events are required to pass the following selection cuts for 300 fb$^{-1}$ at 50 pile up:
\begin{eqnarray}
\textrm{Leading non-$b$-jet $\eta$}& \qquad \left|\eta_{\ell}\right|&\geq 3.0, \nonumber \\
\textrm{Top Mass}& \qquad  160\,{\rm GeV} \leq M_{top} &\leq 180\,{\rm GeV},  \nonumber \\
\textrm{Top P$_{\rm T}$}& \qquad  T_{pt}&\geq 70\,{\rm GeV}\,  \nonumber \\
\textrm{Top Polarization Optimal Basis}& \qquad  T_{pol}&\geq 0\,  \nonumber \\
\textrm{Scalar $H_T$}& \qquad  H_T&\leq 300\,{\rm GeV}\,  \nonumber \\
\label{eq:basiccut}
\end{eqnarray}

For the 3000~fb$^{-1}$ scenario at 14~TeV, the same selection cuts are used except the $H_T$
cut, which is loosened to $H_T \leq 360\,{\rm GeV}$. 


In the case of 33~TeV, the preselection is slightly modified increasing the lepton $p_T$ cut
from 40 to 50~GeV and increasing the jet $p_T$ cut from 70 to 75~GeV. 
Moreover the final selection includes modifications of the same cuts as well as the addition
of leading non-$b$-jet $p_T \geq 100$~GeV. The $H_T$ cut has been loosened to 550~GeV, the top
polarization cut is tightened to 0.2, and the top $p_T$ cut has been tightened to 100~GeV.
These differences select a kinematic region with the highest signal to background ratio
while maintaining sensable cuts based on expected $t$-channel kinematics.

\begin{figure}[!h!tbp]
  \centering

  \subfigure[]{
    \includegraphics[width=0.40\textwidth]{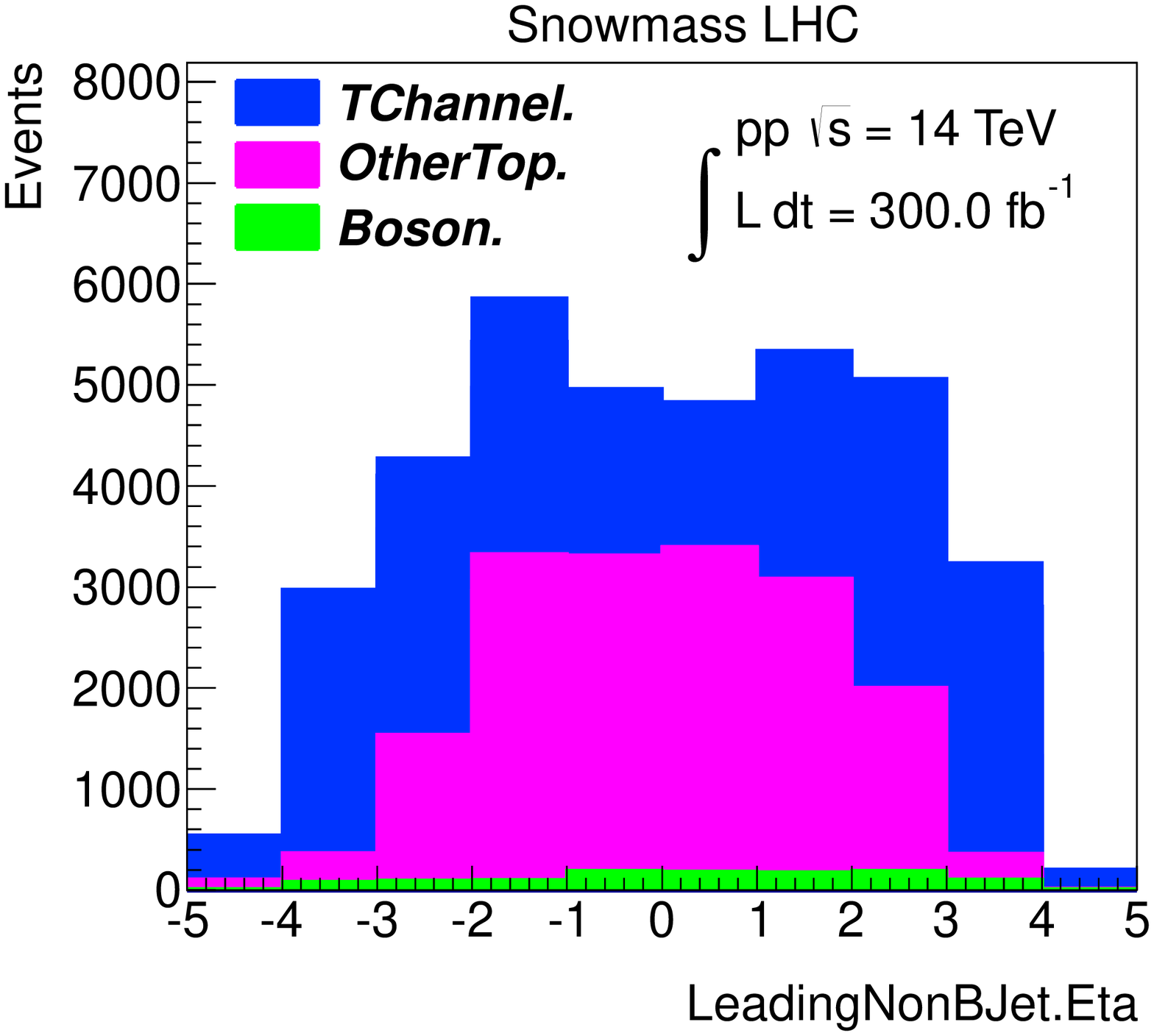}
    \label{fig:a}
  }
  \hspace*{0.0\textwidth}
  \subfigure[]{
    \includegraphics[width=0.40\textwidth]{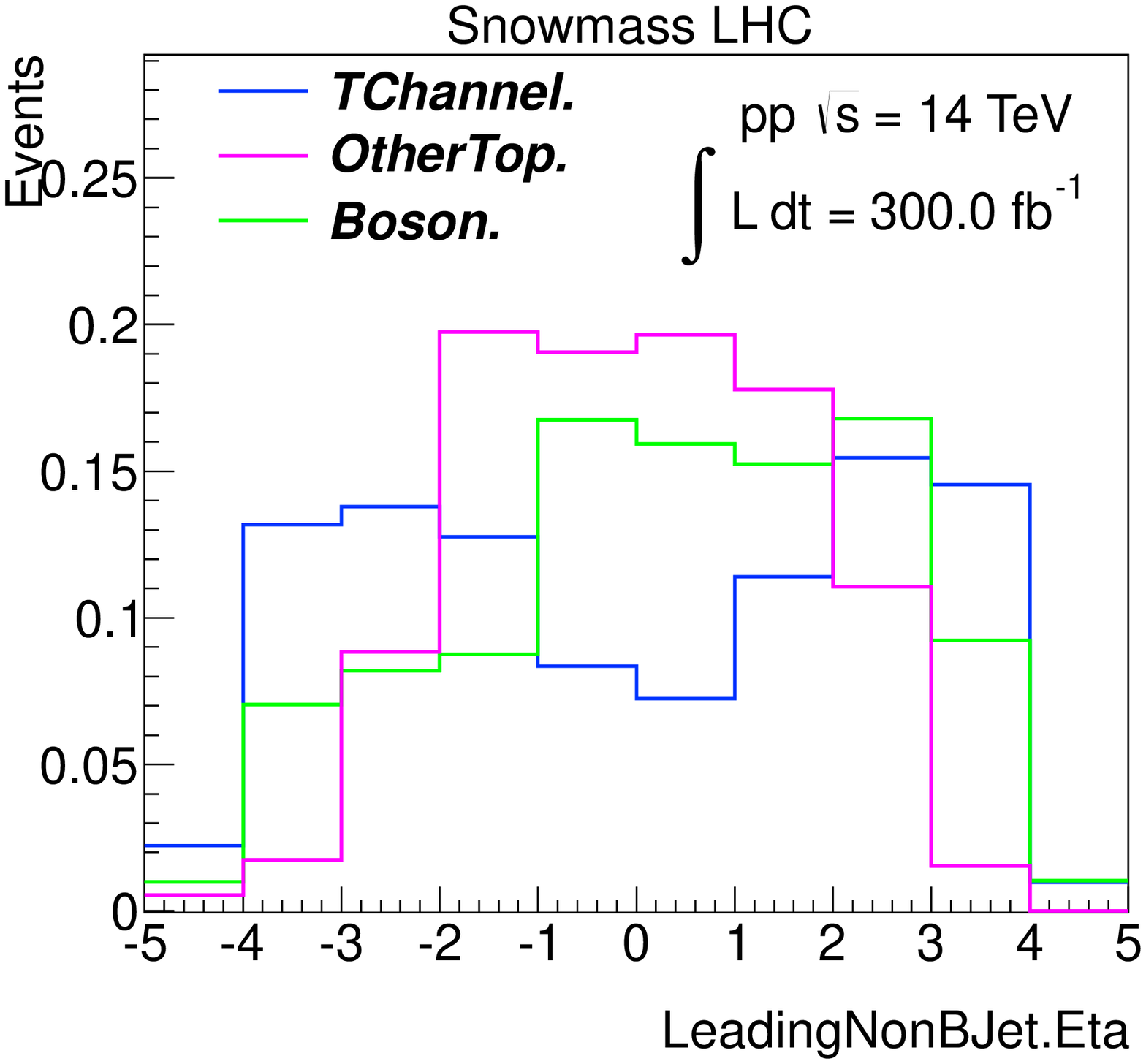}
    \label{fig:b}
  } 
  \vspace*{-0.025\textwidth}

  \subfigure[]{
    \includegraphics[width=0.40\textwidth]{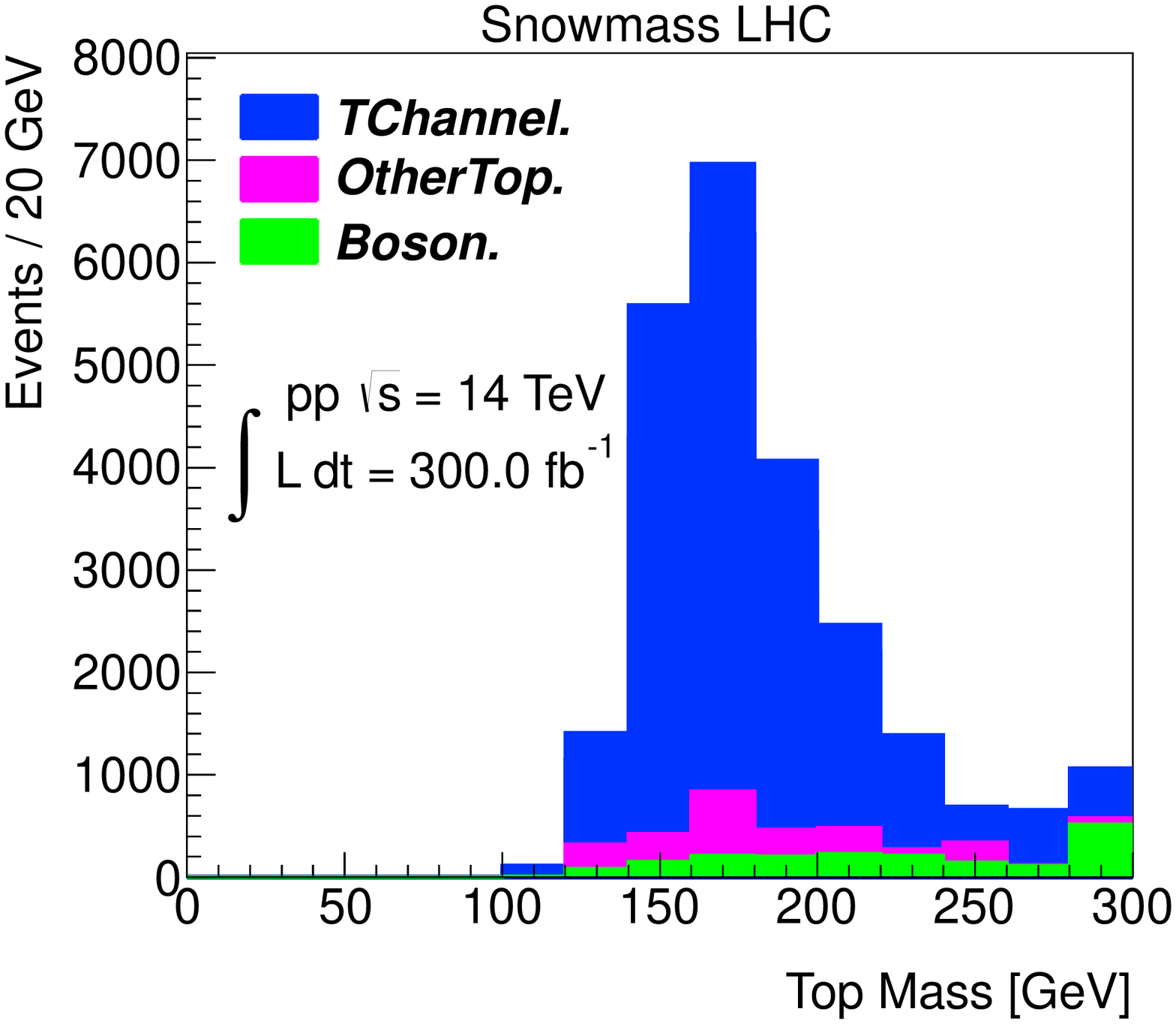}
    \label{fig:c}
  }
  \hspace*{0.0\textwidth}
  \subfigure[]{
    \includegraphics[width=0.40\textwidth]{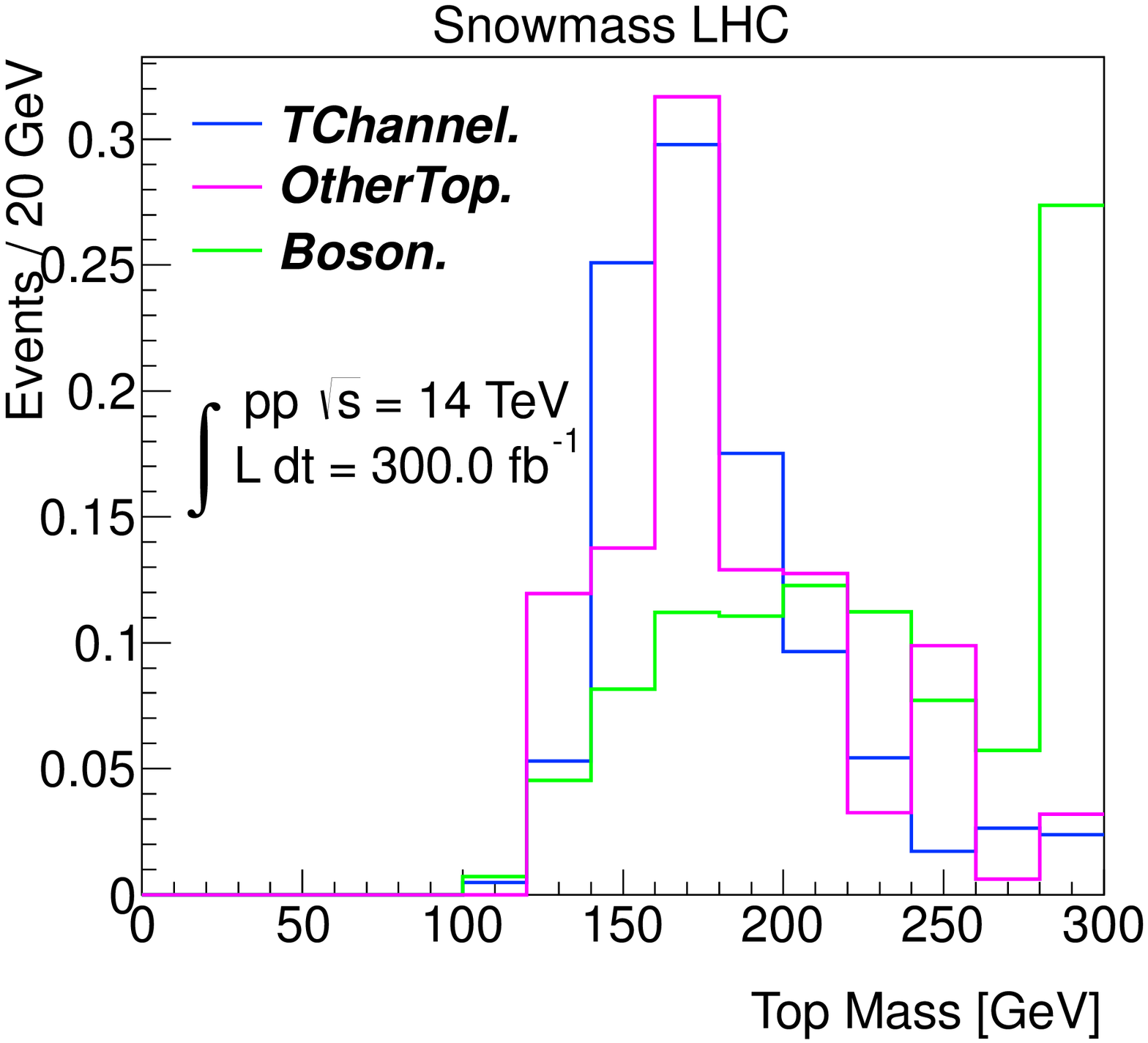}
    \label{fig:d}
  } 
  \vspace*{-0.025\textwidth}

  \subfigure[]{
    \includegraphics[width=0.40\textwidth]{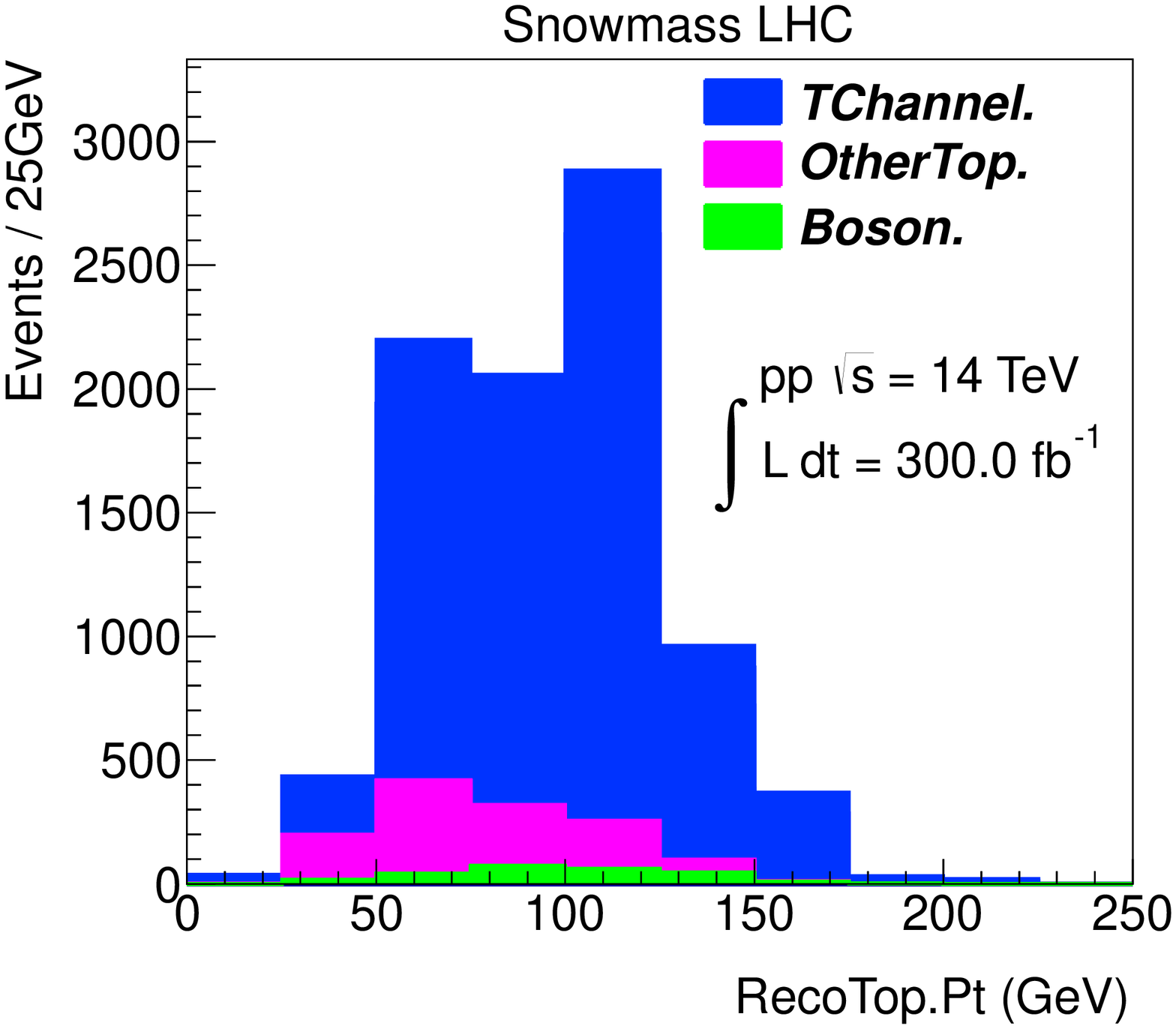}
    \label{fig:e}
  }
  \hspace*{0.0\textwidth}
  \subfigure[]{
    \includegraphics[width=0.40\textwidth]{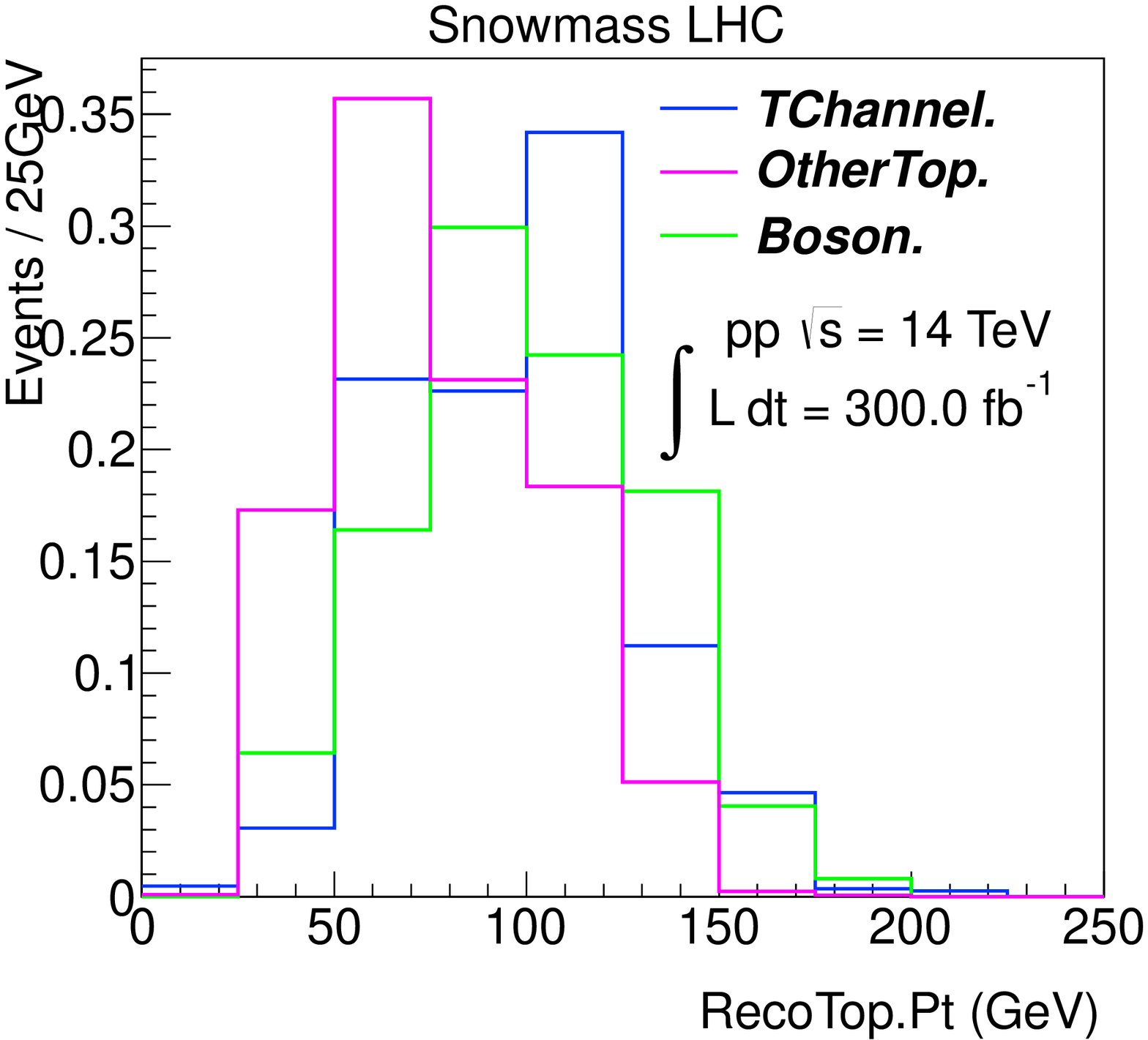}
    \label{fig:f}
  } 
  \caption{Discriminant variables after all cuts have been applied except on the variable
shown for 300~fb$^{-1}$ at 14~TeV: (a, b) $\eta$ of the leading non-$b$-jet, (c, d)  $p_T$ of
the leading non-$b$-jet, and (e, f) mass of the top quark. Figures~(a), (c) and (e) are
normalized to the expected event yield while (b), (d) and (f) are normalized to unit area.}
  \label{fig:kinematics}
\end{figure}

\begin{figure}[!h!tbp]
  \centering
  
  \subfigure[]{
    \includegraphics[width=0.40\textwidth]{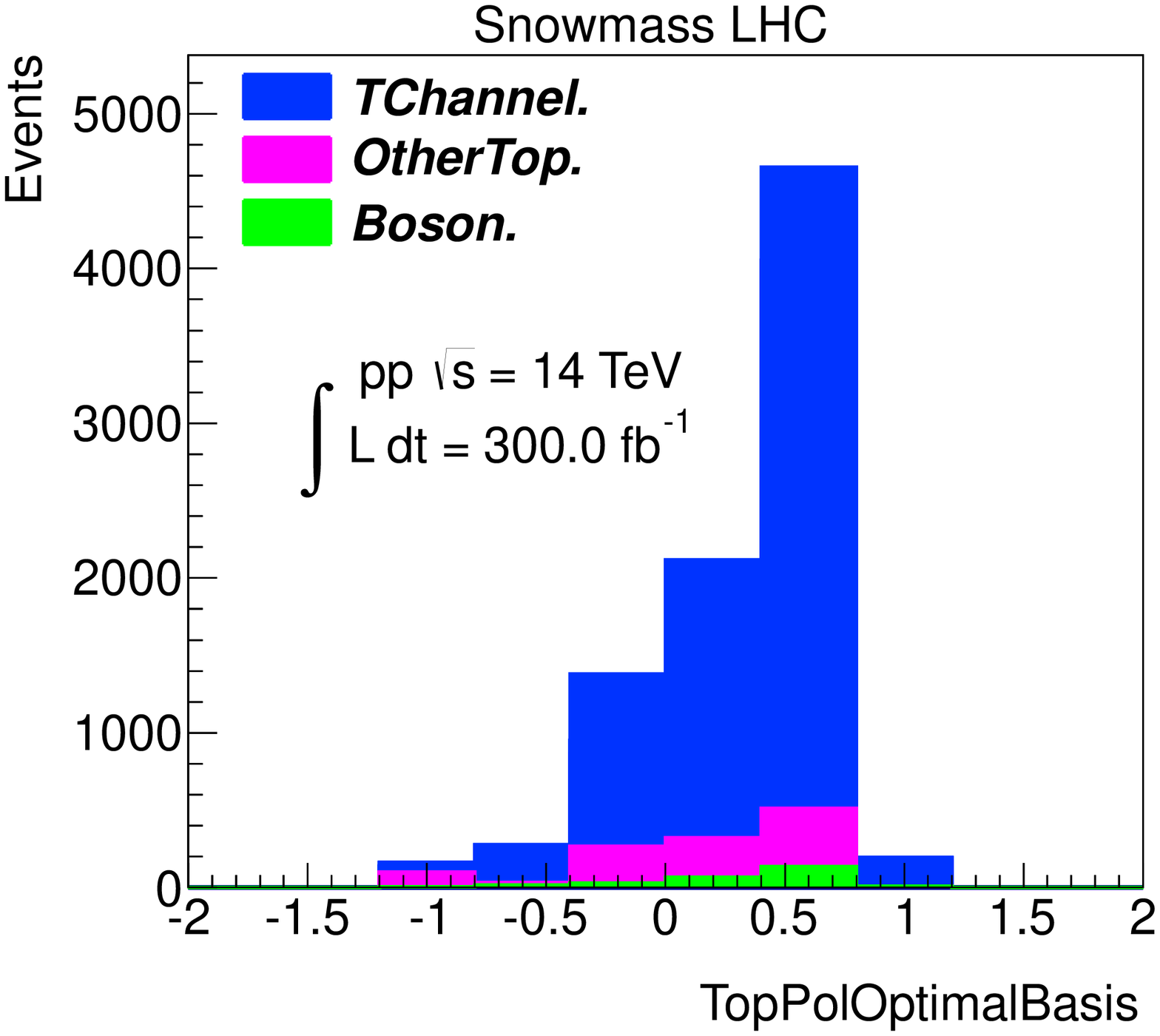}
    \label{fig:a}
  }
  \hspace*{0.0\textwidth}
  \subfigure[]{
    \includegraphics[width=0.40\textwidth]{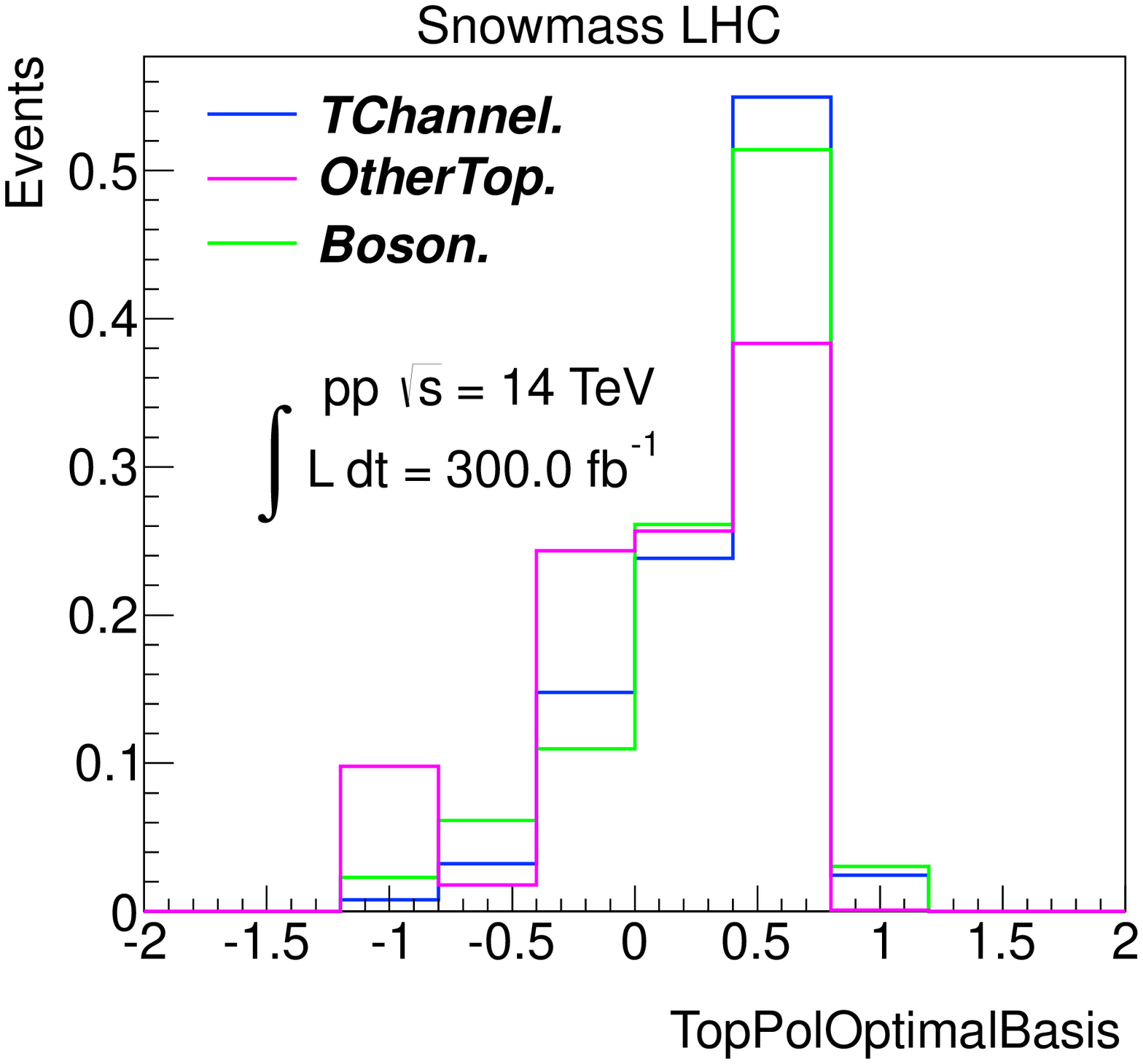}
    \label{fig:b}
  } 
  
  \subfigure[]{
    \includegraphics[width=0.40\textwidth]{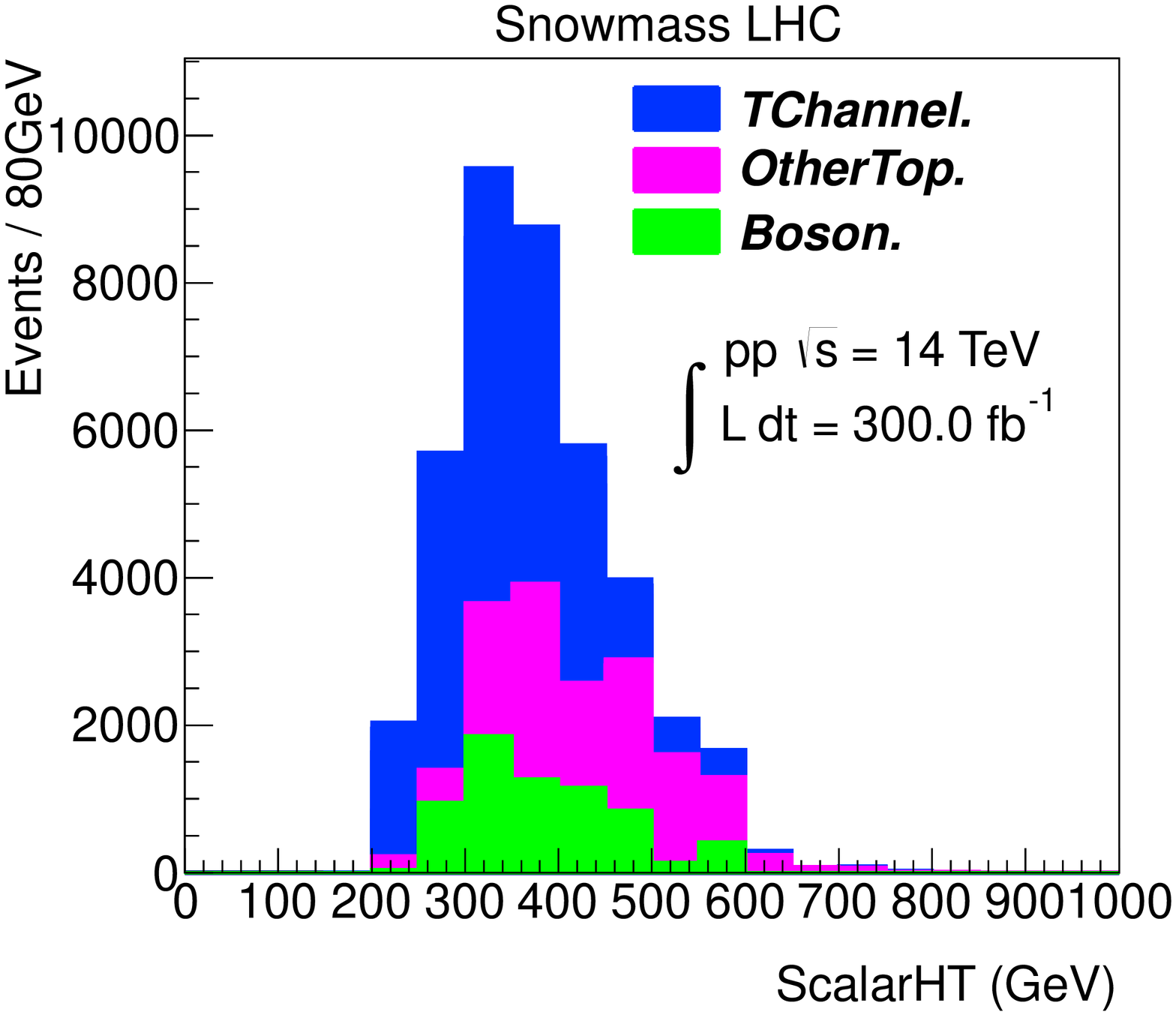}
    \label{fig:c}
  }
  \hspace*{0.0\textwidth}
  \subfigure[]{
    \includegraphics[width=0.40\textwidth]{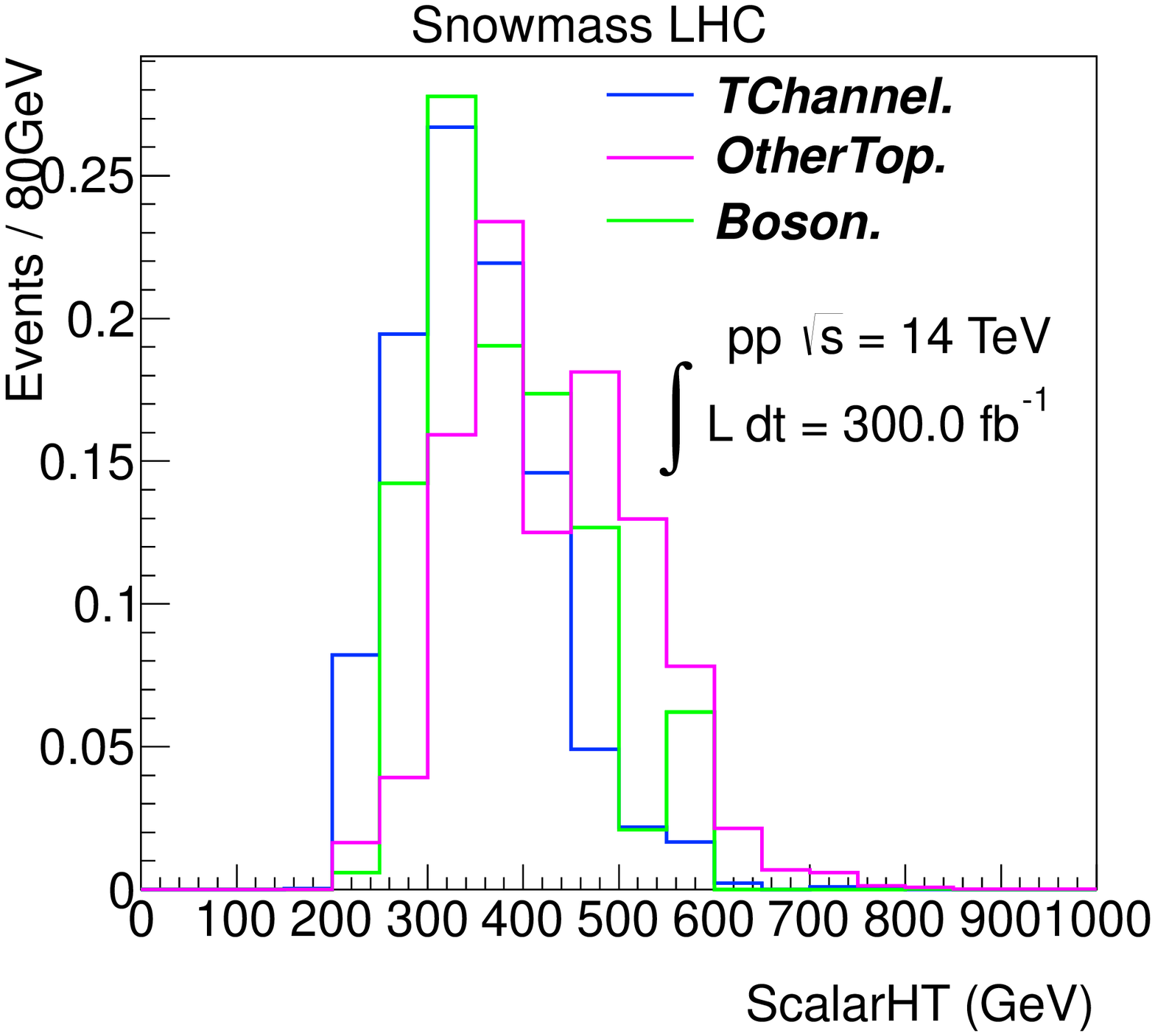}
    \label{fig:d}
  } 
  
  \caption{Discriminant variables after all cuts have been applied except on the variable
shown for 300~fb$^{-1}$ at 14~TeV: (a, b) $p_T$ of the top quark ,(c, d) polarization of the
top quark, and (e, f) scalar sum of all $p_T$ objects. Figures~(a) and (c) are normalized
to the expected event yield while (b) and (d) are normalized to unit area.}
  \label{fig:kinematics2}
\end{figure}

\begin{figure}[!h!tbp]
  \centering

  \subfigure[]{
    \includegraphics[width=0.40\textwidth]{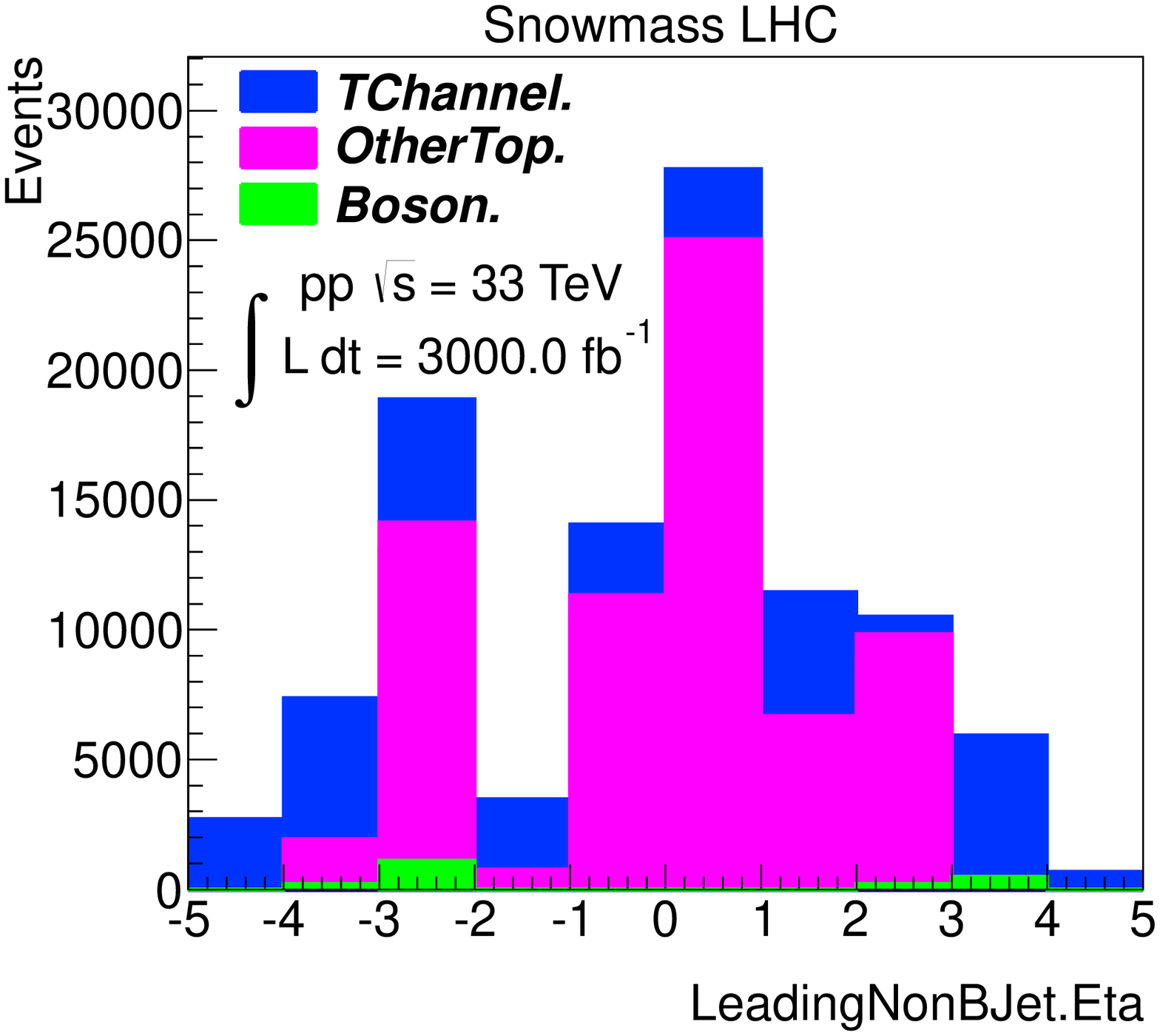}
    \label{fig:a}
  }
  \hspace*{0.0\textwidth}
  \subfigure[]{
    \includegraphics[width=0.40\textwidth]{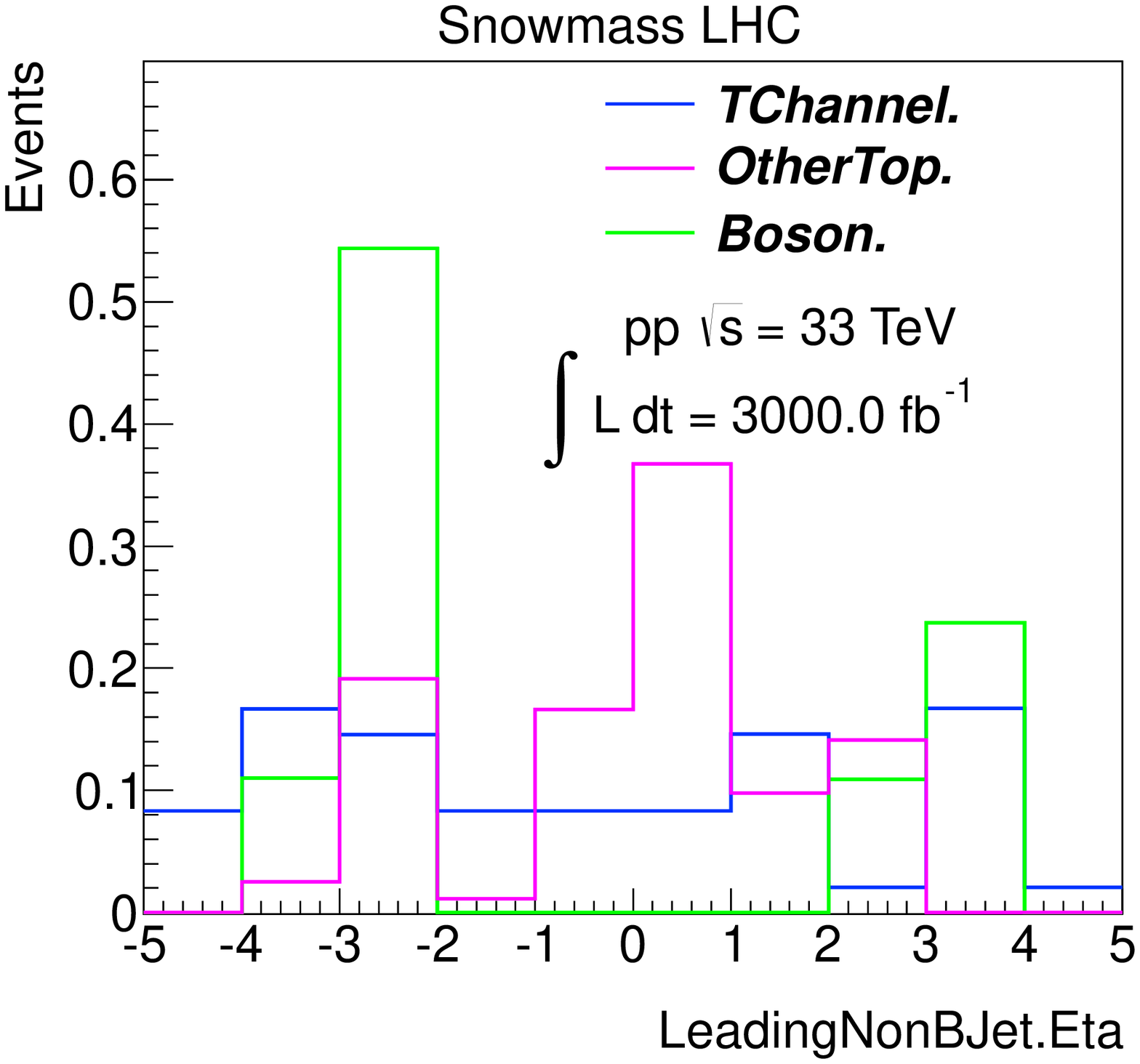}
    \label{fig:b}
  } 
  \vspace*{-0.025\textwidth}

 \subfigure[]{
    \includegraphics[width=0.40\textwidth]{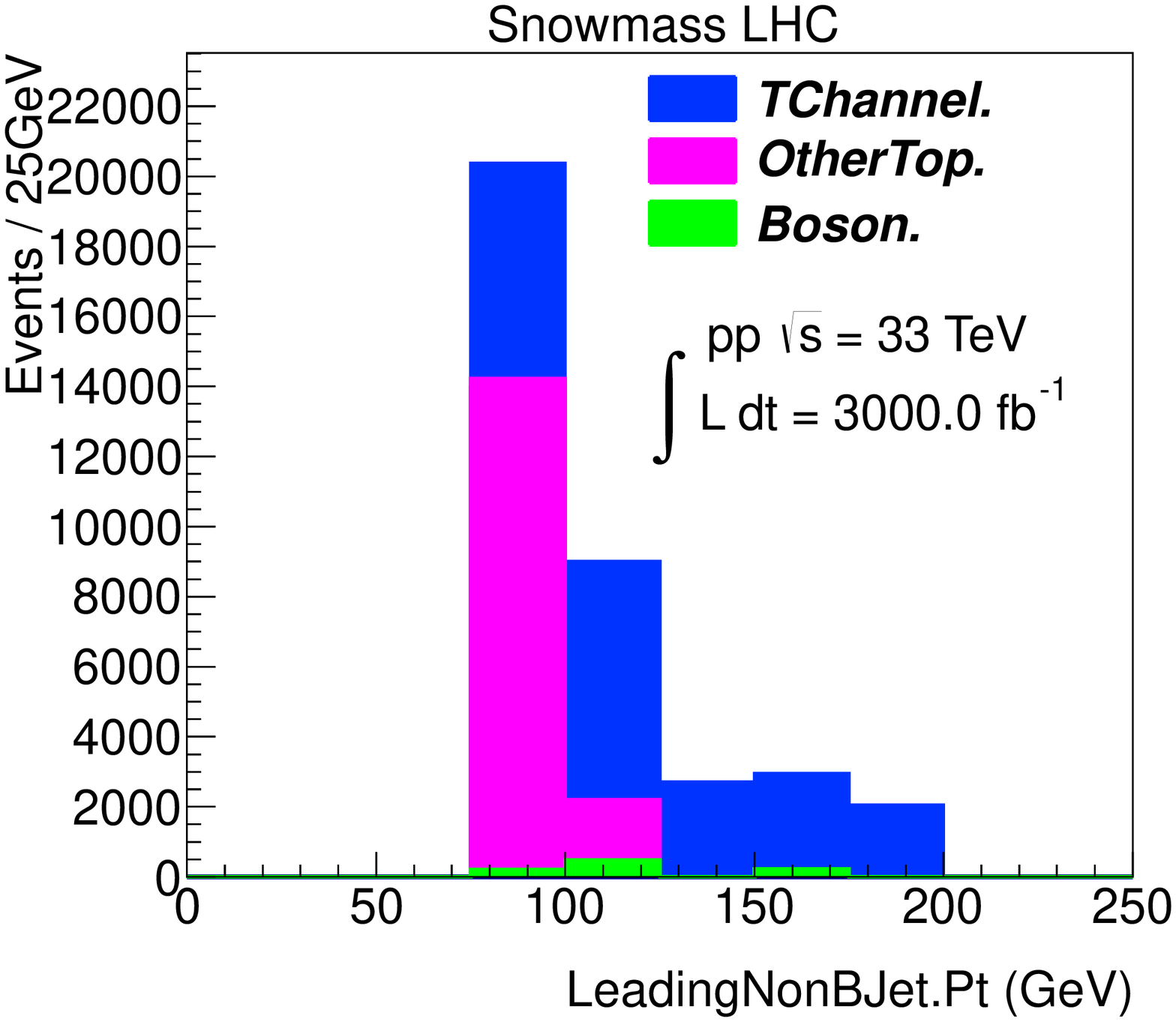}
    \label{fig:c}
  }
  \hspace*{0.0\textwidth}
  \subfigure[]{
    \includegraphics[width=0.40\textwidth]{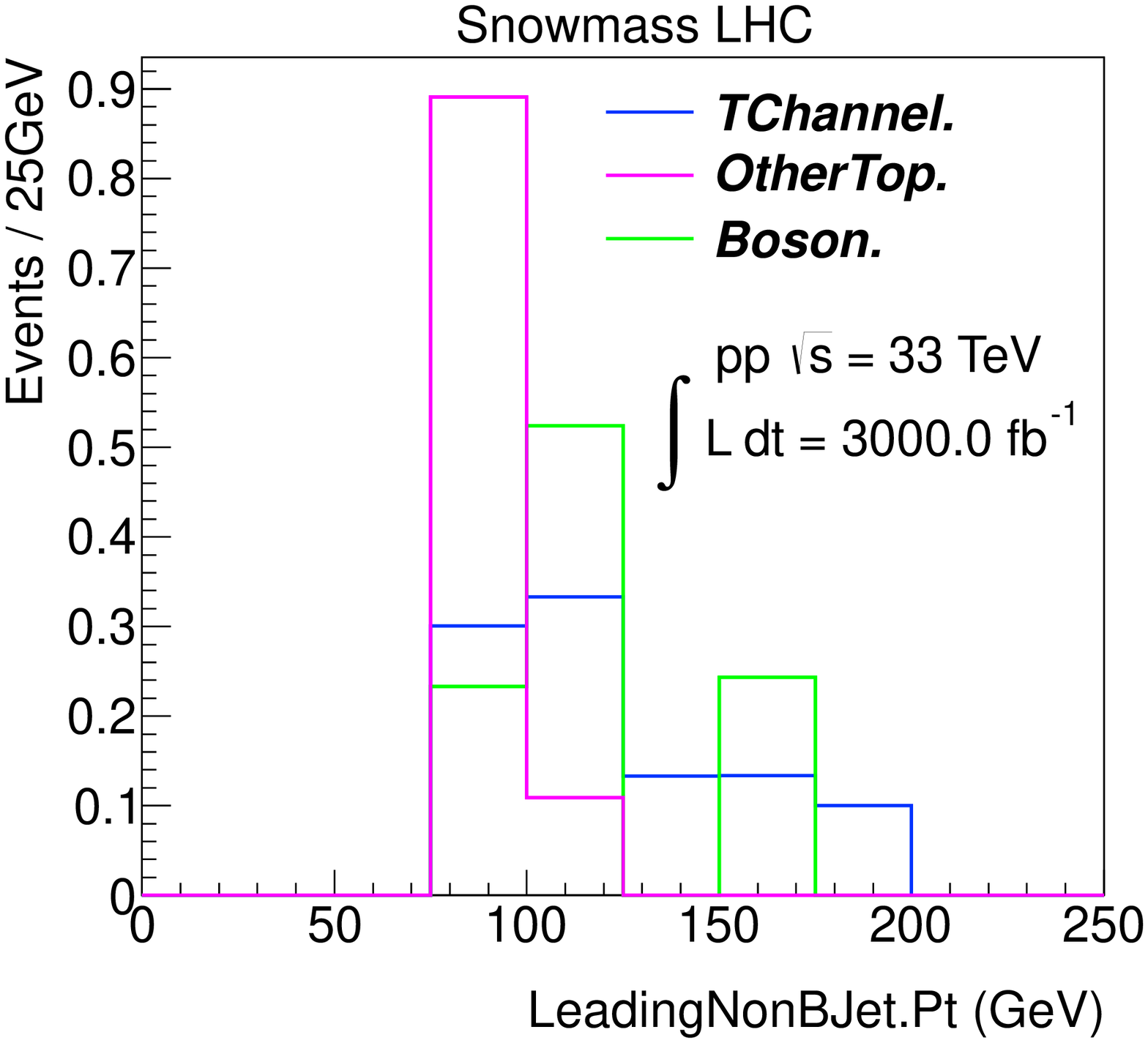}
    \label{fig:d}
  } 
  \vspace*{-0.025\textwidth}

  \subfigure[]{
    \includegraphics[width=0.40\textwidth]{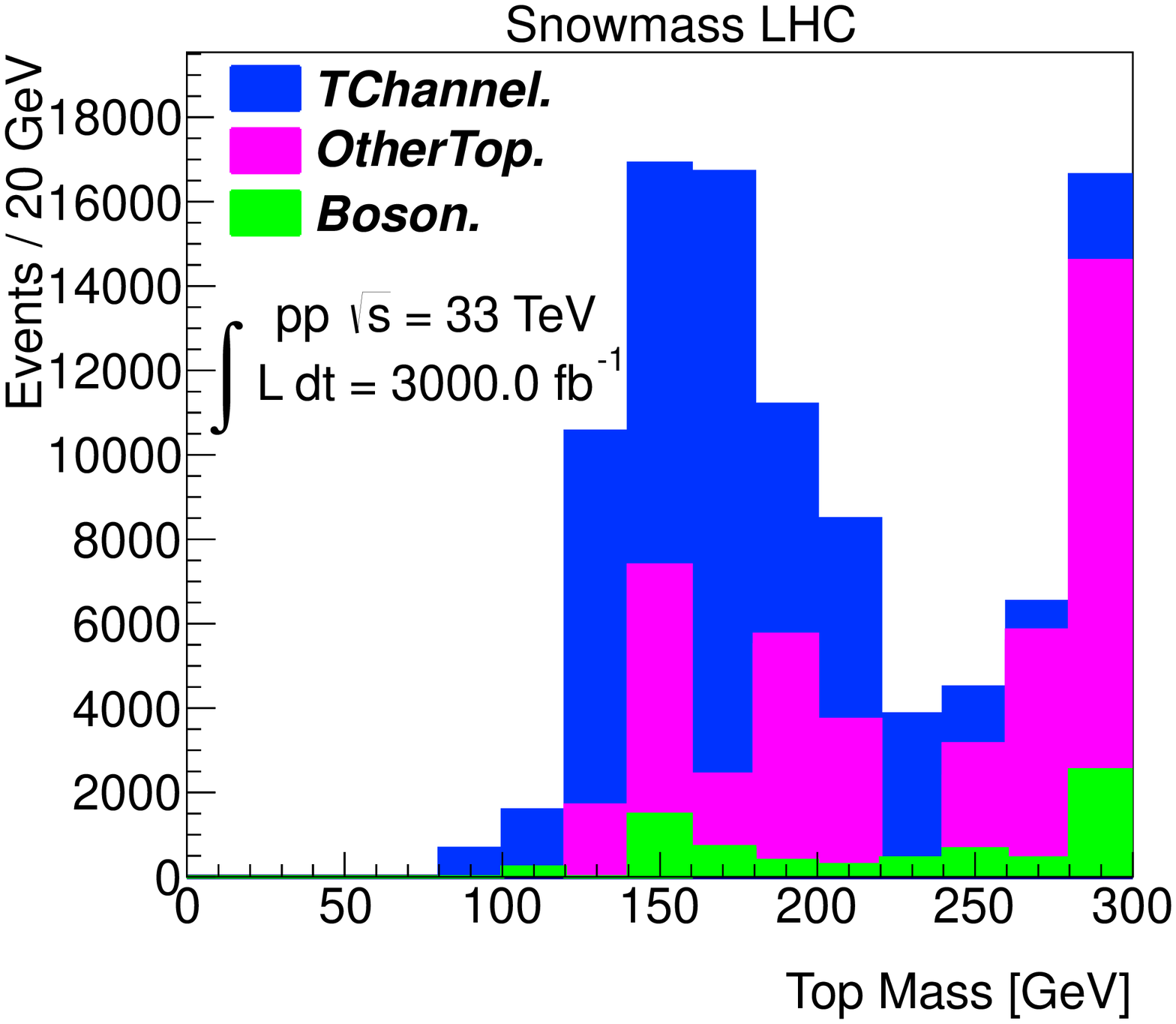}
    \label{fig:e}
  }
  \hspace*{0.0\textwidth}
  \subfigure[]{
    \includegraphics[width=0.40\textwidth]{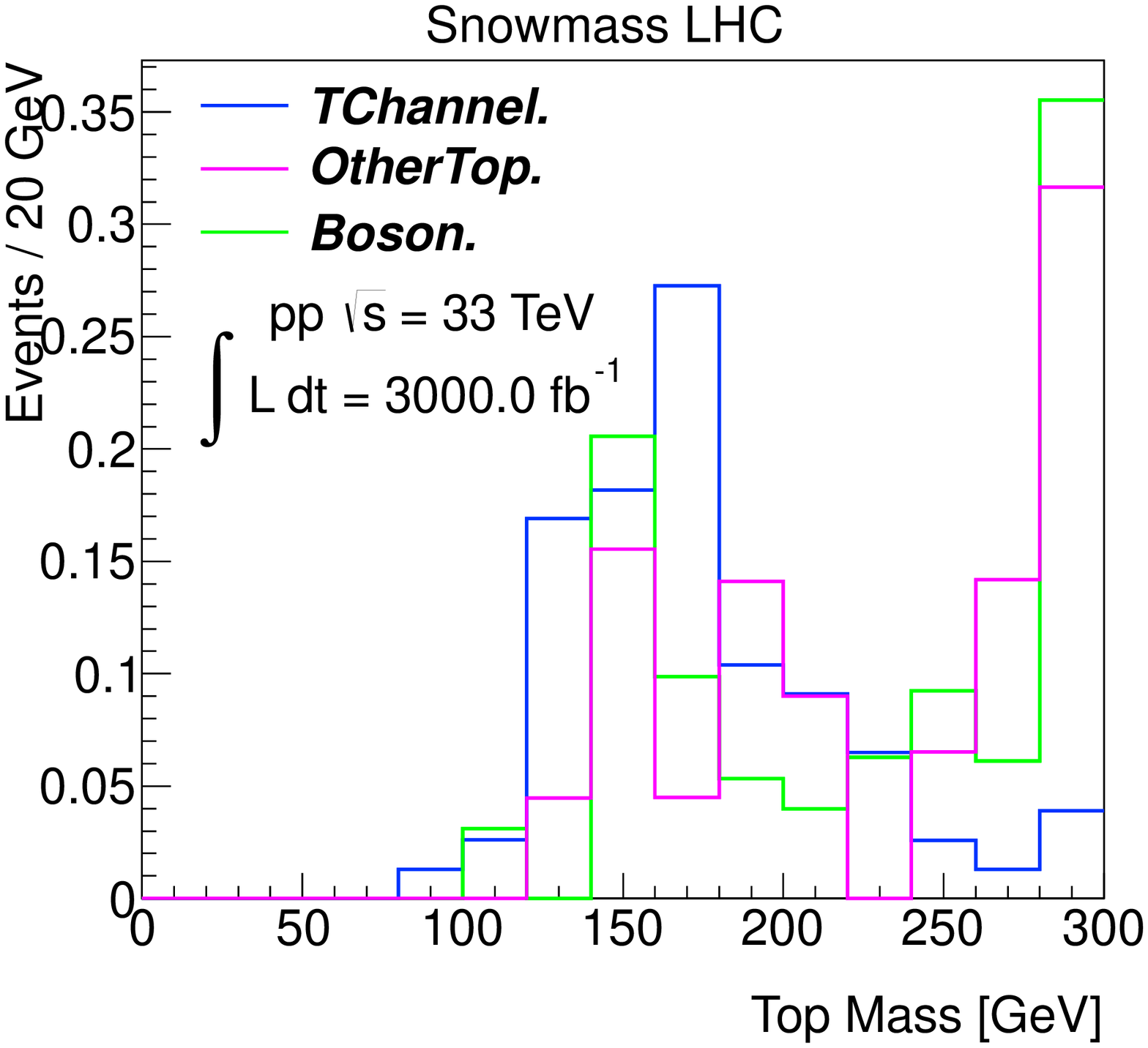}
    \label{fig:f}
  }

  \caption{Discriminant variables after all cuts have been applied except on the variable
shown for 3000~fb$^{-1}$ at 33~TeV: (a, b) $\eta$ of the leading non-$b$-jet, (c, d)  $p_T$ of
the leading non-$b$-jet, and (e, f) mass of the top quark. Figures~(a), (c) and (e) are
normalized to the expected event yield while (b), (d) and (f) are normalized to unit area.}
  \label{fig:kinematics33}
\end{figure}

\begin{figure}[!h!tbp]
  \centering

  \subfigure[]{
    \includegraphics[width=0.40\textwidth]{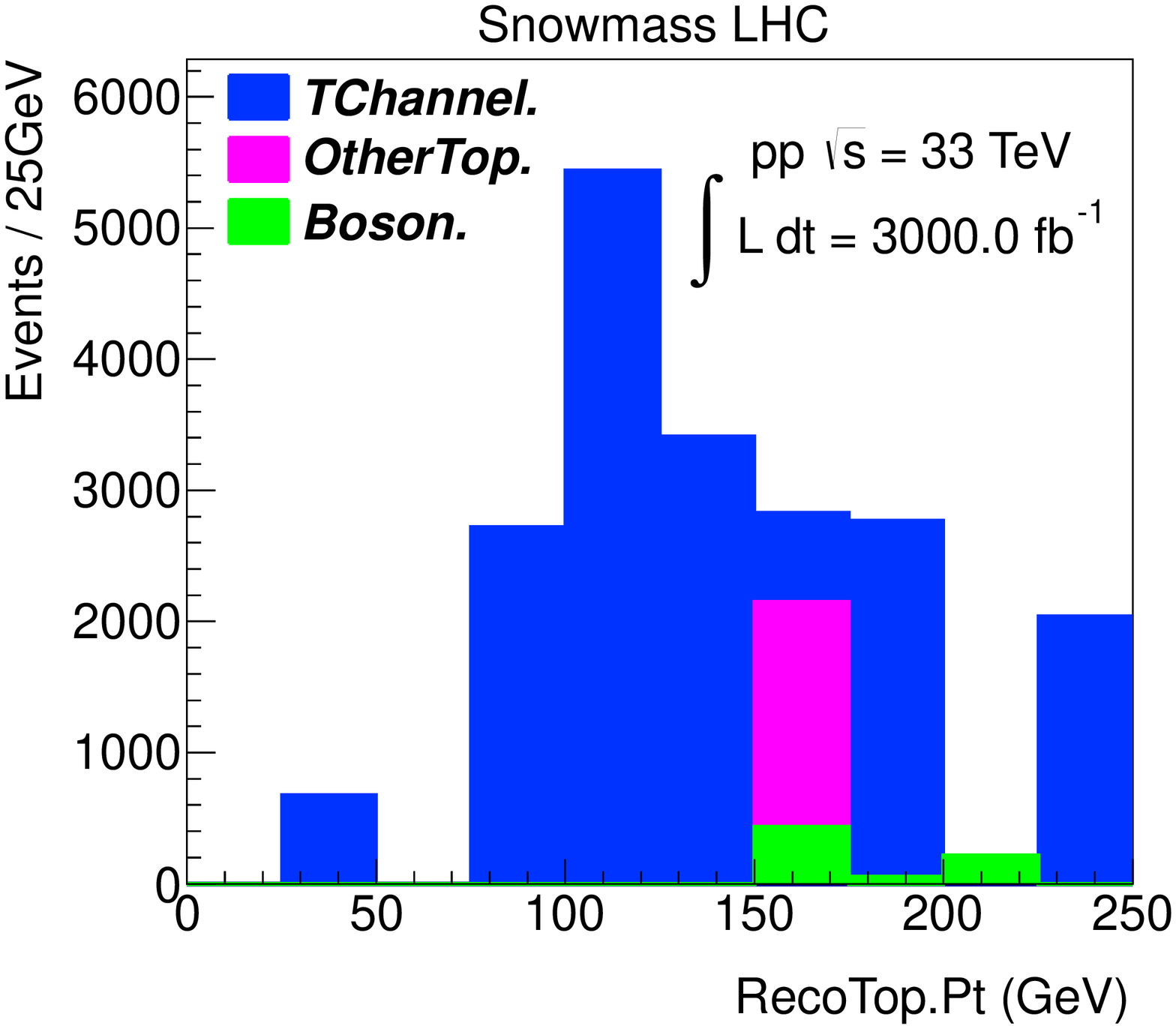}
    \label{fig:e}
  }
  \hspace*{0.0\textwidth}
  \subfigure[]{
    \includegraphics[width=0.40\textwidth]{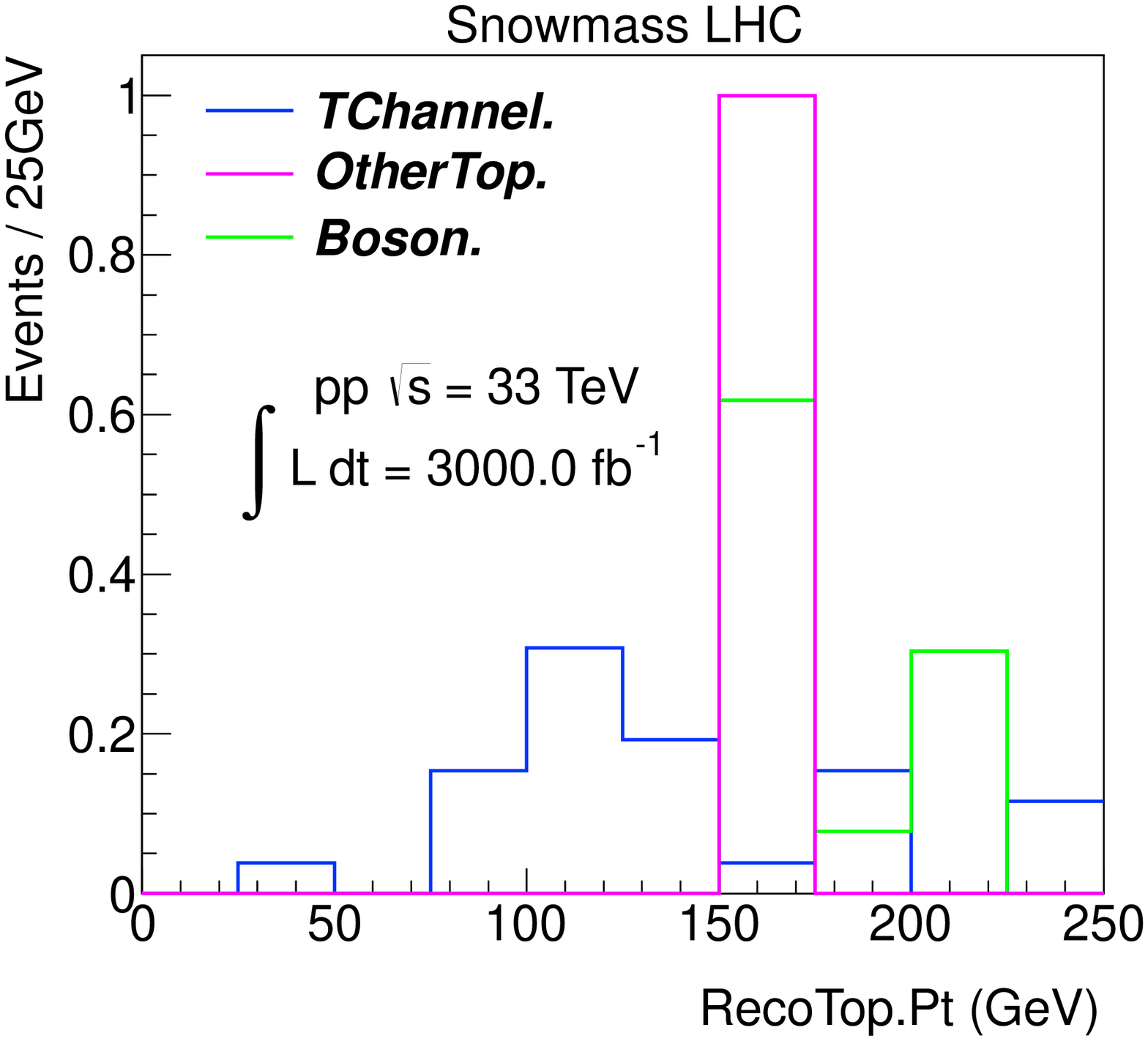}
    \label{fig:f}
  }   
  \vspace*{-0.025\textwidth}

  \subfigure[]{
    \includegraphics[width=0.40\textwidth]{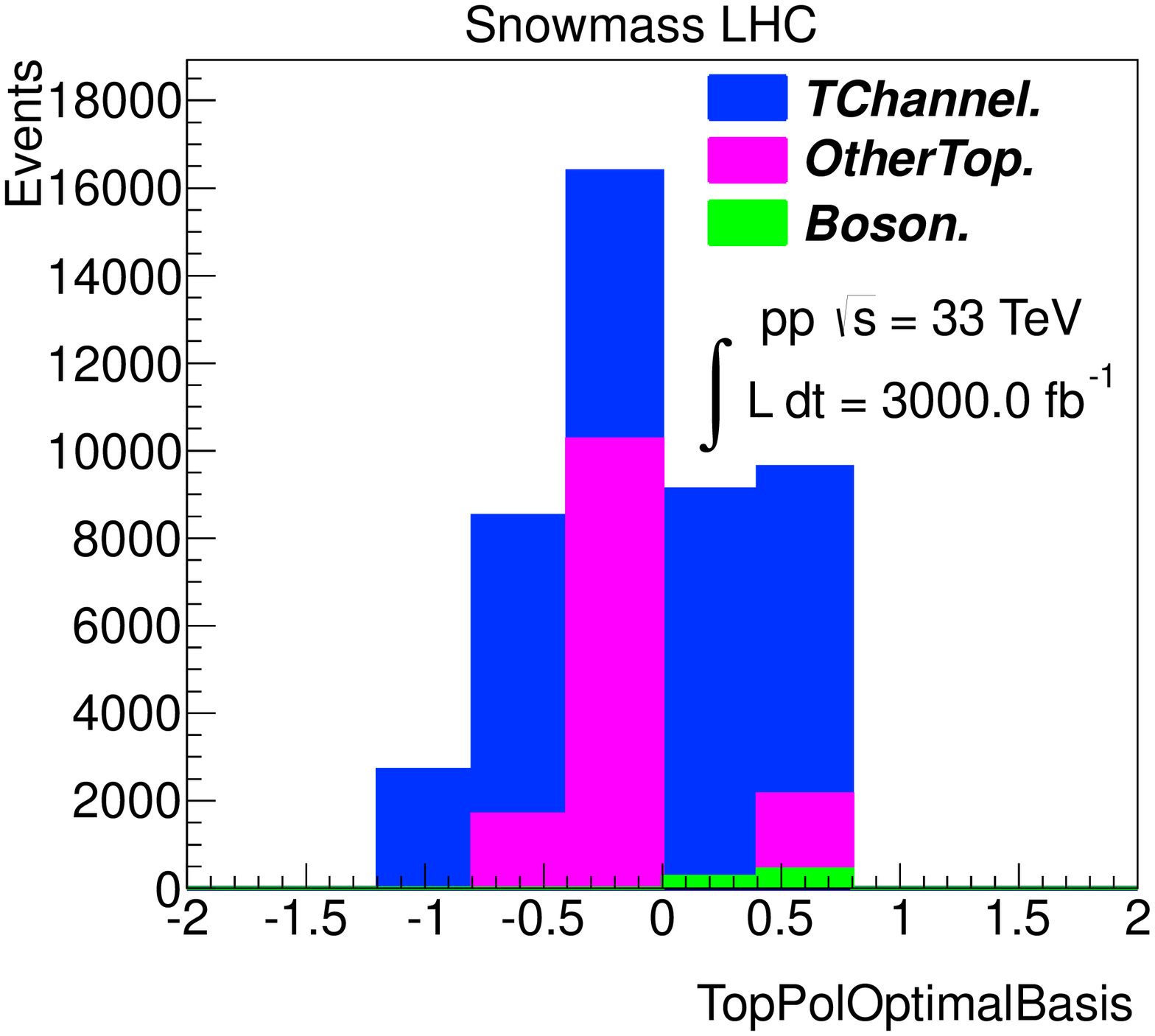}
    \label{fig:a}
  }
  \hspace*{0.0\textwidth}
  \subfigure[]{
    \includegraphics[width=0.40\textwidth]{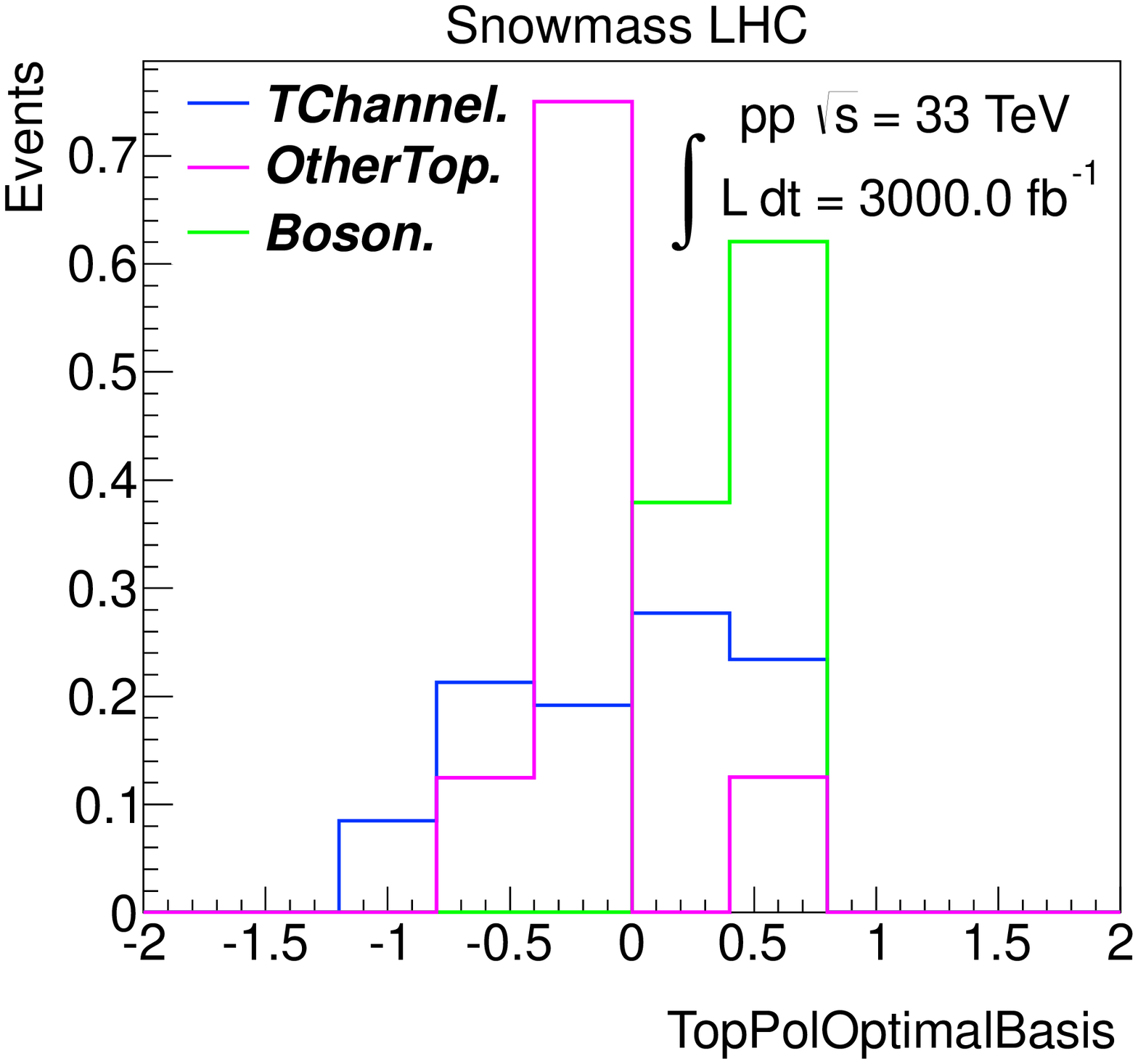}
    \label{fig:b}
  } 
  \vspace*{-0.025\textwidth}

  \subfigure[]{
    \includegraphics[width=0.40\textwidth]{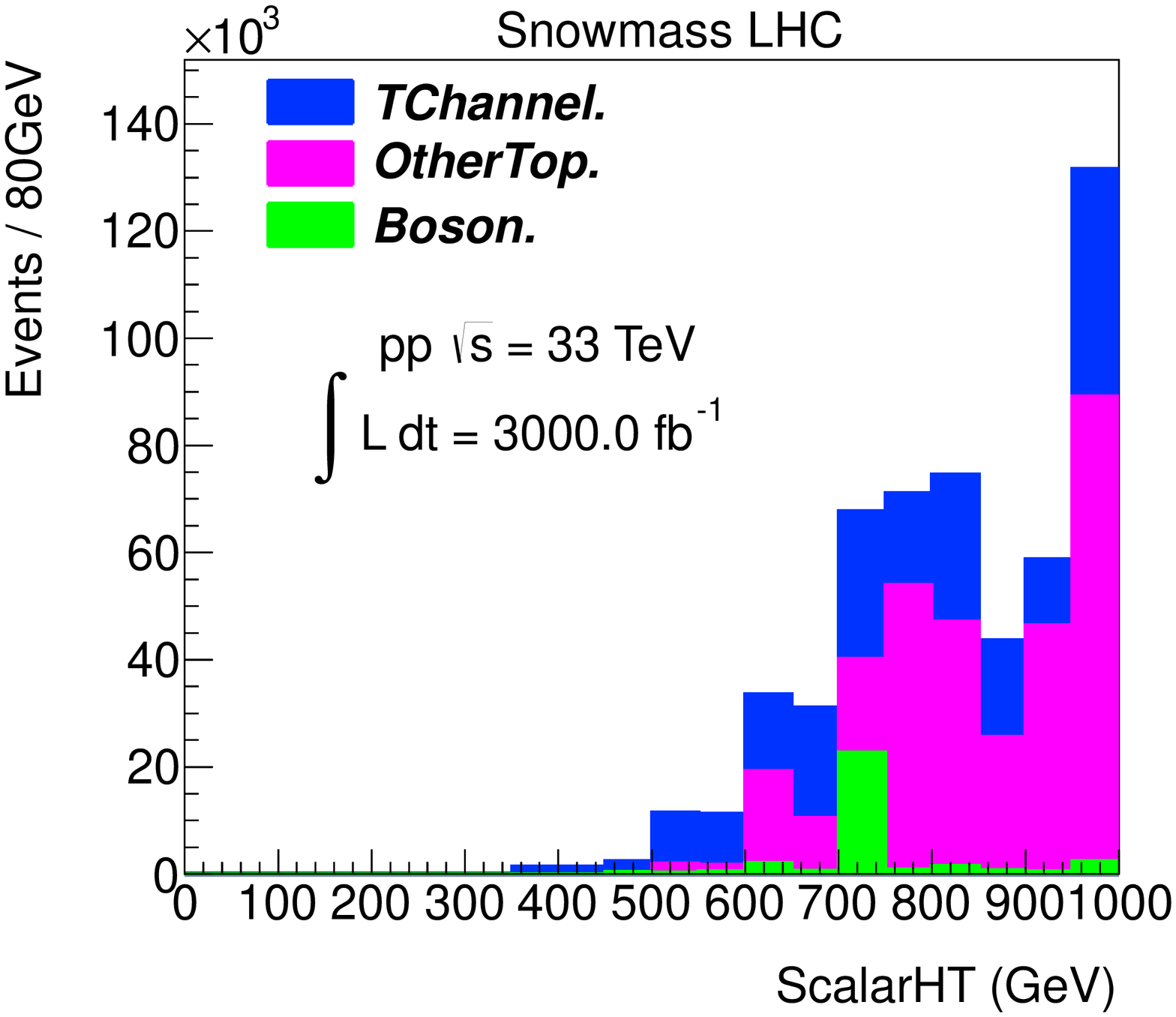}
    \label{fig:c}
  }
  \hspace*{0.0\textwidth}
  \subfigure[]{
    \includegraphics[width=0.40\textwidth]{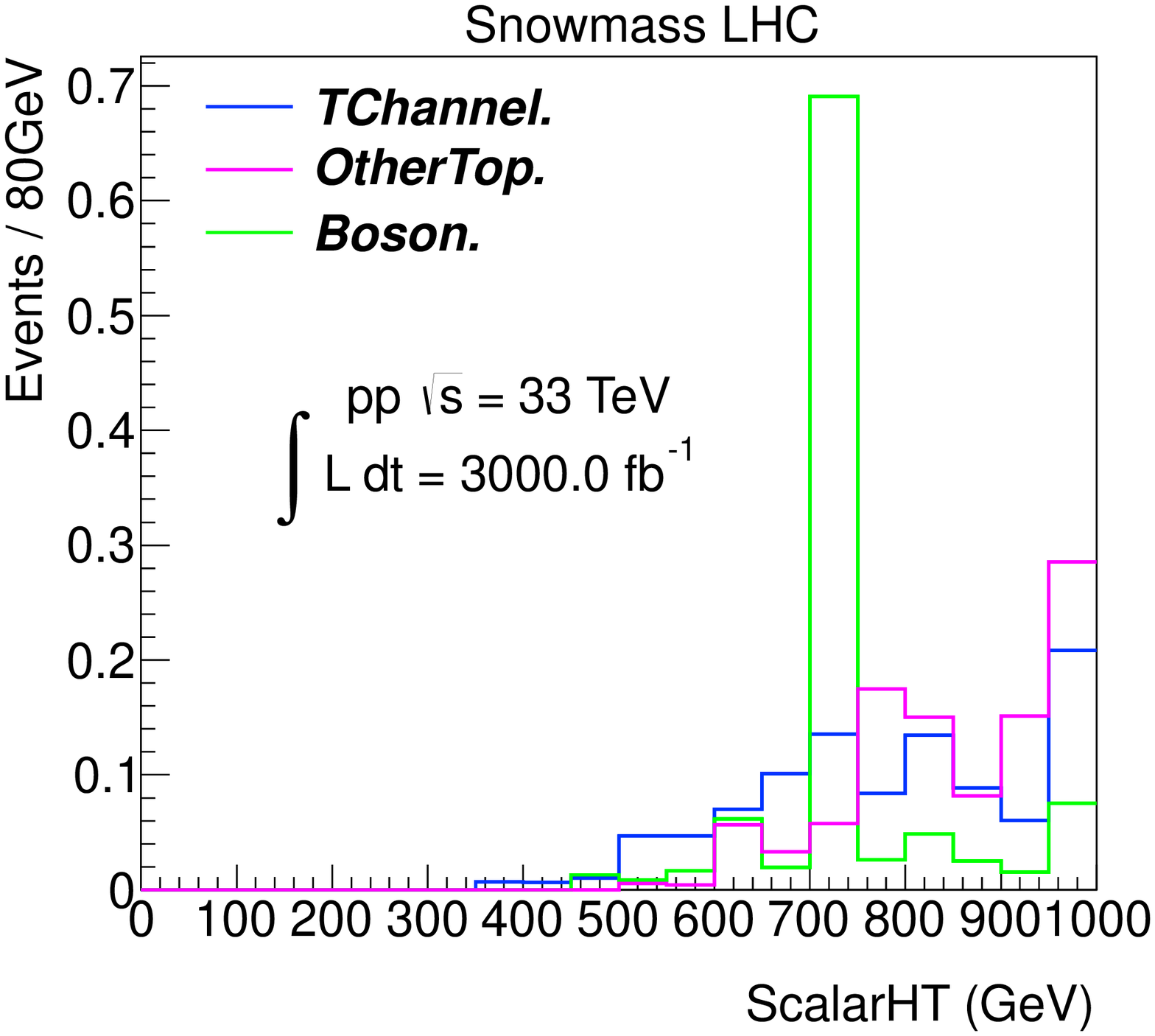}
    \label{fig:d}
  } 
  
  \caption{Discriminant variables after all cuts have been applied except on the variable
shown for 3000~fb$^{-1}$ at 33~TeV: (a, b) $p_T$ of the top quark ,(c, d) polarization of the
top quark, and (e, f) scalar sum of all $p_T$ objects. Figures~(a) and (c) are normalized
to the expected event yield while (b) and (d) are normalized to unit area.}
  \label{fig:kinematics33_2}
\end{figure}

Figures~\ref{fig:kinematics} and~\ref{fig:kinematics2} show the effect of the cuts for the
300~fb$^{-1}$ 14~TeV sample. For each histogram, all cuts are applied except on the variable
shown. 
Figures~\ref{fig:kinematics33} and~\ref{fig:kinematics33_2} show the same for 33~TeV.

AT 14~TeV it is clearly visible that the most powerful variables to cut on are the leading
non-$b$-jet $\eta$ and $H_T$,
whereas the top mass, $p_T$ and polarization only add some additional discrimination. At
33~TeV, the distributions are broader and the available statistics are lower, but the 
leading non-$b$-jet $\eta$ and $p_T$ are powerful variables again, whereas now the separation
power of $H_T$ is limited.

Tables~\ref{tab:yields50}, ~\ref{tab:yields140} and~\ref{tab:yields33} show the number of
events passing each cut for 50 pile up at 14~TeV, 140 pile up at 14~TeV, and 140 pile up at
33~TeV respectively. The signal contribution over the total background after all cuts
is about 7.3 for 50 pile up at 14~TeV, 8.2 for 140 pile up at 14~TeV, and 5.9 for 140 pile up
at 33~TeV. The number of expected $t$-channel signal events is 6000, 9000, and 14000,
respectively. These are large, clean event samples suitable for further single top studies.

\begin{table}[!h!tbp]
  \caption{Cutflow of signal and background yields in 300~fb$^{-1}$ at 14~TeV.}
  \label{tab:yields50}
  \begin{center}
    \renewcommand{\arraystretch}{1.4}
    \begin{tabular}{|l|c|c|c|c|c|c|}
      \hline
      Event Yield & Signal  &  $Wt$  & $t\bar{t}$ & diboson & $W+\rm{jets}$ & Total Background  \\ 
      \hline
      no cuts                    & 1.81e+07  & 2.10e+07   &   2.14e+08   &   1.09e+08    &  3.62e+07   &    3.80e+08 \\
      \hline
      $+$~preselection             &  746000	&    439000    &  5.01e+06    &  122000   &   102000  &	5.67e+06	  \\ 
      \hline
      $+$~Leading non-$b$-jet $\eta$   &    159000  	&    13900    &    187000   &   3850  &  20200  &  226000     \\
      \hline
      $+$~Top mas window          &   33600    &  1460    	&   25700   &  105  &   1060  & 28300 \\
      \hline
      $+$~Top $p_T$          &  29300   &   1270    	&   22100    &   97.8  &   964   &    24400\\ 
      \hline
      $+$~ Top Polarization      & 22100   &  700           &   10600    &   74.3   & 769   &   12100 \\
      \hline
      $+$~ $H_T$            &   6120   &   57.4          &    570    &   12.5  &  197  &  836 \\
      \hline
            
    \end{tabular}
  \end{center}
\end{table}

\begin{table}[!h!tbp]
  \caption{Cutflow of signal and background yields in 3000~fb$^{-1}$ at 14~TeV.}
  \label{tab:yields140}
  \begin{center}
    \renewcommand{\arraystretch}{1.4}
    \begin{tabular}{|l|c|c|c|c|c|c|}
      \hline
      Event Yield & Signal  &  $Wt$  & $t\bar{t}$ & diboson & $W+\rm{jets}$ & Total Background  \\ 
      \hline
      no cuts                    & 1.81e+08  &   2.10e+08       &  2.13e+09    &  1.09e+09     &  3.62e+08   &   3.79e+09  \\
      \hline
      $+$~preselection             & 7.46e+06 &    4.23e+06     &  4.96e+07    &   1.22e+06   &   1.04e+06   &	5.61e+07	  \\ 
      \hline
      $+$~Leading non-$b$-jet $\eta$   &   1.68e+06 &   189000     &    2.57e+06    &   52800   &  201000   &   3.01e+06    \\
      \hline
      $+$~Top mas window          &  306000     &   17400   	&  274000    &  2580  &  9970   & 304000 \\
      \hline
      $+$~Top $p_T$                  & 263000    &  14300    	&  229000     &   2530   &   8870    & 255000   \\ 
      \hline
      $+$~ Top Polarization      & 189000   &   8180          &  125000     &    481   & 6980   &  140000  \\
      \hline
      $+$~ $H_T$                   & 9380     &     115        &  672       &   0.0   &  345  &  1130 \\
      \hline
            
    \end{tabular}
  \end{center}
\end{table}

\begin{table}[!h!tbp]
  \caption{Cutflow of signal and background yields in 3000~fb$^{-1}$ at 33~TeV.}
  \label{tab:yields33}
  \begin{center}
    \renewcommand{\arraystretch}{1.4}
    \begin{tabular}{|l|c|c|c|c|c|c|}
      \hline
      Event Yield & Signal  &  $Wt$  & $t\bar{t}$ & diboson & $W+\rm{jets}$ & Total Background  \\ 
      \hline
      no cuts                    & 2.52e+09  & 9.88e+08   &  1.13e+10    &   2.70e+09    &  8.28e+08   &  1.59e+10   \\
      \hline
      $+$~preselection             &  1.66e+07 &   1.25e+07     & 1.60e+08     &  4.30e+06   &    2.04e+06  &  1.79e+08		  \\ 
      \hline
      $+$~Leading non-$b$-jet $\eta$   &    5.25e+06  &   1.51e+06     &   1.98e+07     &  458000   & 585000   &   2.23e+07    \\
      \hline
      $+$~Leading non-$b$-jet $p_T$   &   3.74e+06  &   557000     &   8.13e+06    &   190000  &   426000  &  9.30e+06     \\
      \hline
      $+$~top mas window          &  666000     &  47000    	&  1.00e+06    &  10000  &  16800   &  1.07e+06\\
      \hline
      $+$~top Pt                  &   595000   &  33000     	&    842000    &  9200   &  14300    & 898000   \\ 
      \hline
      $+$~ Top Polarization      & 184000   &    8700         &   237000    &   2250   &   7970  &  256000  \\
      \hline
      $+$~ $H_T$             &   14200    &    0.0         &  1710      &   54.0  &  650  &  2410 \\
      \hline
            
    \end{tabular}
  \end{center}
\end{table}

A $t$-channel cross-section measurement can be extracted in a simple analysis by assuming
that the background is estimated with a large systematic uncertainty of 30\% that accounts for
both detector modeling (including jet energy scale and pileup and $b$-tagging) and
theory uncertainties. No systematic uncertainty is assigned to the signal. The resulting
signal uncertainty is then added in quadrature with the statistical uncertainty to give the
total expected uncertainty shown in Table~\ref{tab:xsprec}.

\begin{table}[!h!tbp]
 \begin{center}
    \begin{tabular}{|l|c|c|c|}
      \hline
      Collider    &  statistical      &  systematic  & total  \\
                  &  uncertainty [\%] &  uncertainty [\%]& uncertainty [\%] \\
      \hline
      300fb$^{-1}$, 14~TeV & 1.4      &  4.1         &  4.3 \\
      3000fb$^{-1}$, 14~TeV & 1.1      &  3.6         &  3.8 \\
      3000fb$^{-1}$, 33~TeV & 0.9      &  5.1         &  5.2 \\
      \hline     
     \end{tabular}
    \caption{Estimated $t$-channel cross section precision assuming a 30\% background
systematic uncertainty.}
    \label{tab:xsprec}
 \end{center} 
\end{table}

Even though the relative background uncertainty is large, the signal cross-section can still
be measured with high precision because the selected sample consist almost entirely of
$t$-channel signal events. The signal precision improves at the high-luminosity LHC due
to the larger event sample.

\begin{figure}[!h!tbp]
  \centering
  \subfigure[]{
    \includegraphics[width=0.40\textwidth]{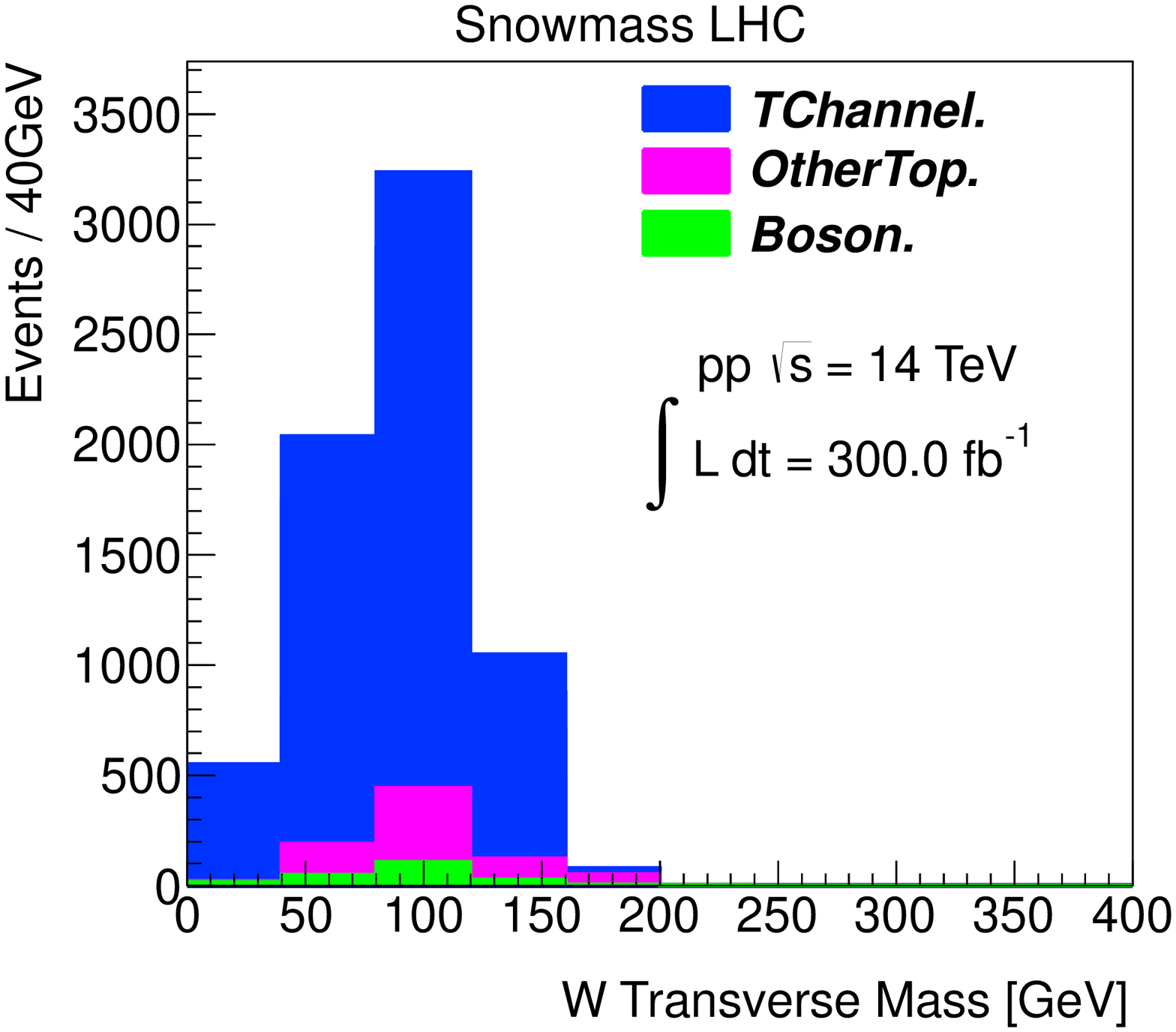}
    \label{fig:a}
  }
  \hspace*{0.0\textwidth}
  \subfigure[]{
    \includegraphics[width=0.40\textwidth]{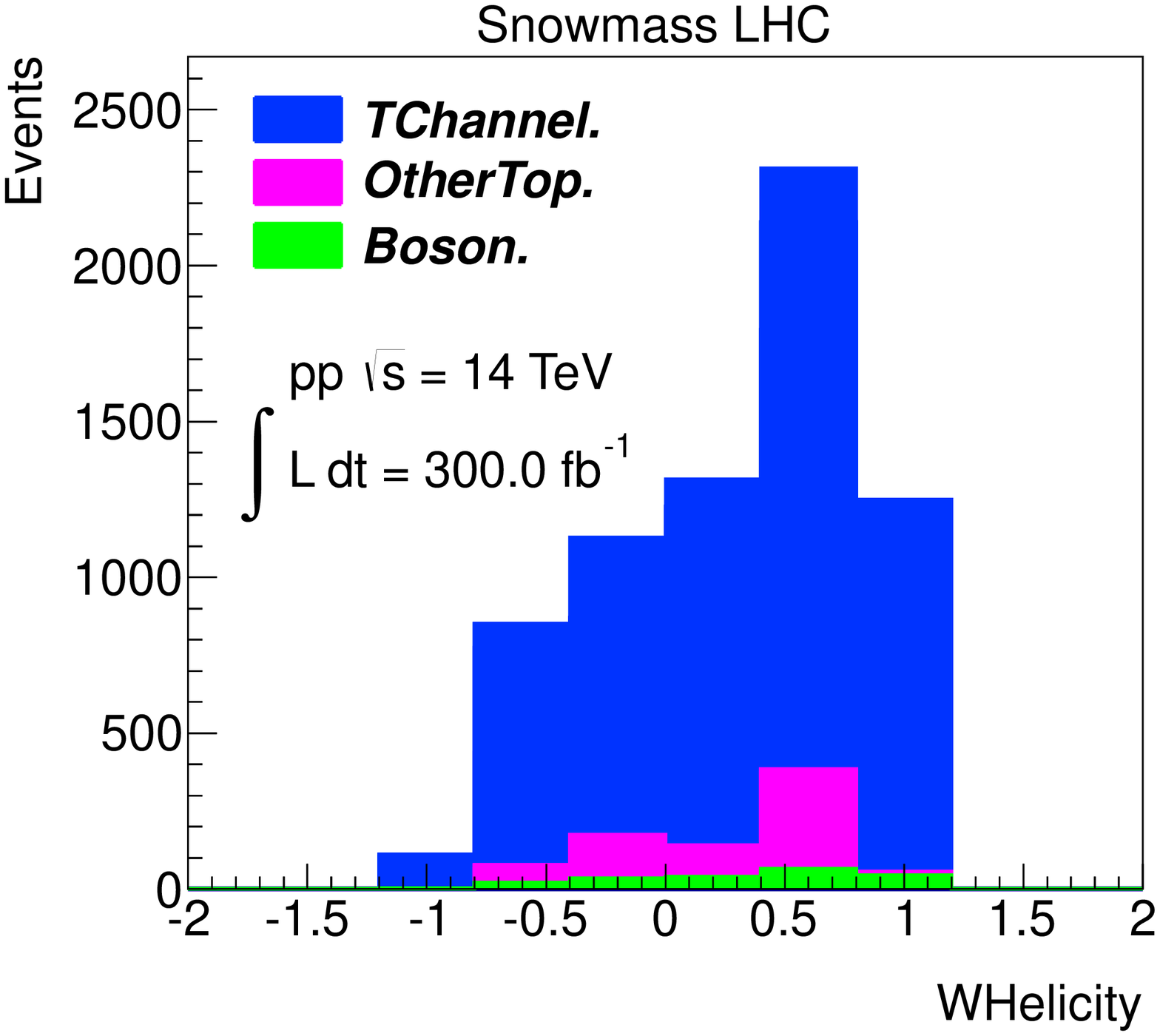}
    \label{fig:b}
  }
  \subfigure[]{
    \includegraphics[width=0.40\textwidth]{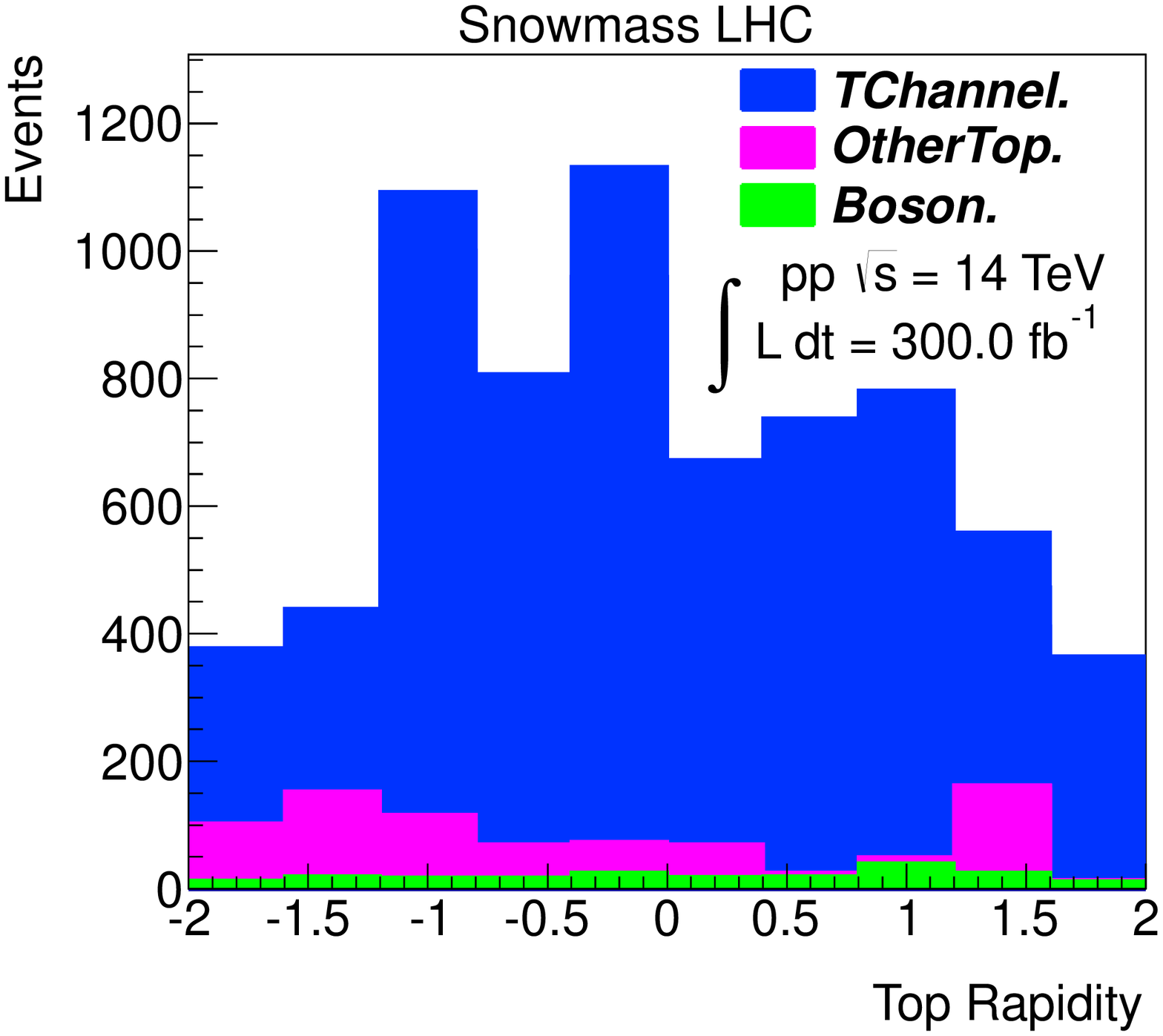}
    \label{fig:c}
  } 
  \hspace*{0.0\textwidth}
  \subfigure[]{
    \includegraphics[width=0.40\textwidth]{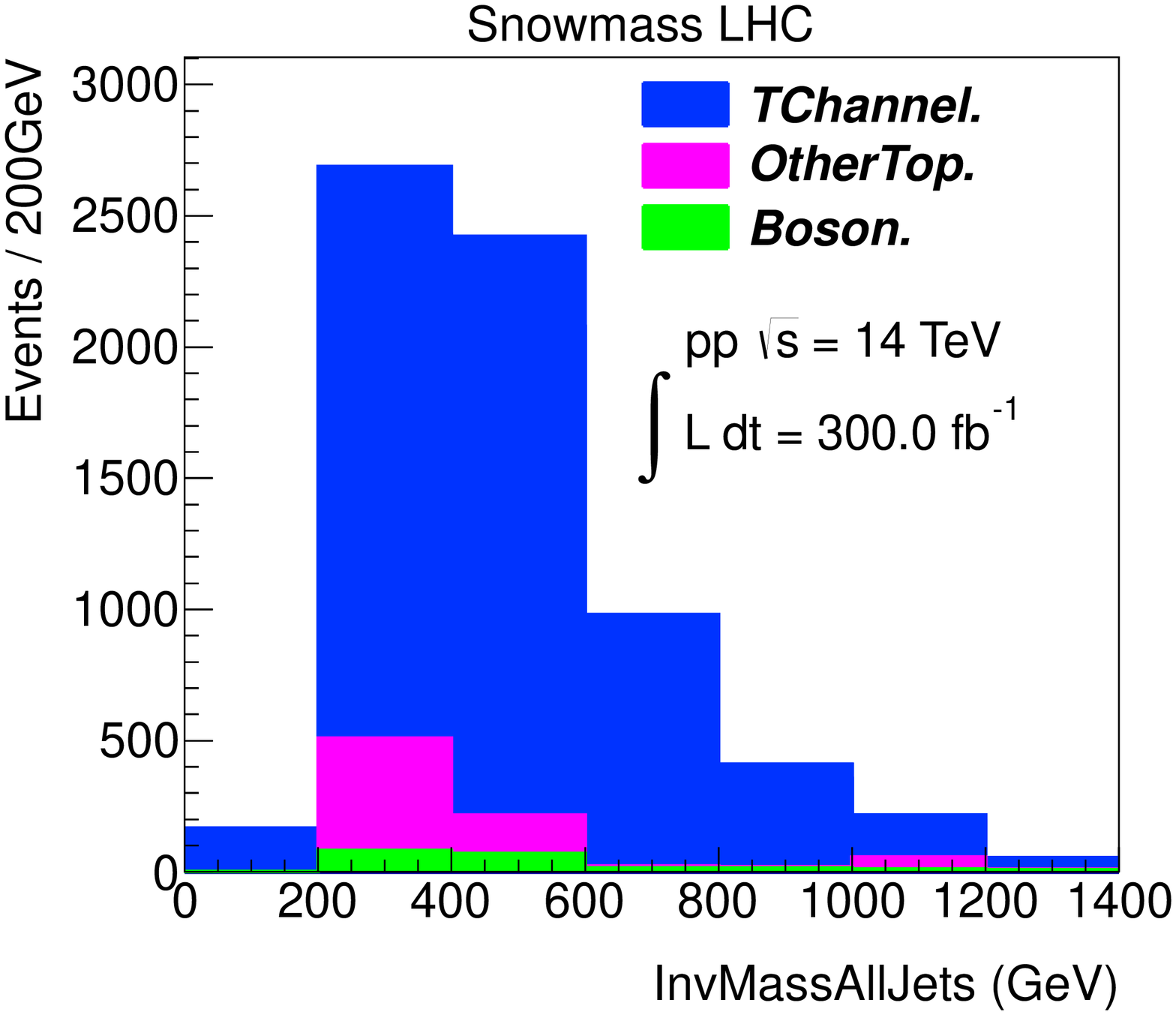}
    \label{fig:d}
  } 
  \caption{Kinematic distributions for events passing selection cuts for 300~fb$^{-1}$
at 14~TeV: (a) transverse mass of the W boson, (b) helicity of the W boson, (c) rapidity of
the top quark  and (d) invariant mass of all the jets in the event.}
  \label{fig:afterkinematics}
\end{figure}

\begin{figure}[!h!tbp]
  \centering
  \subfigure[]{
    \includegraphics[width=0.40\textwidth]{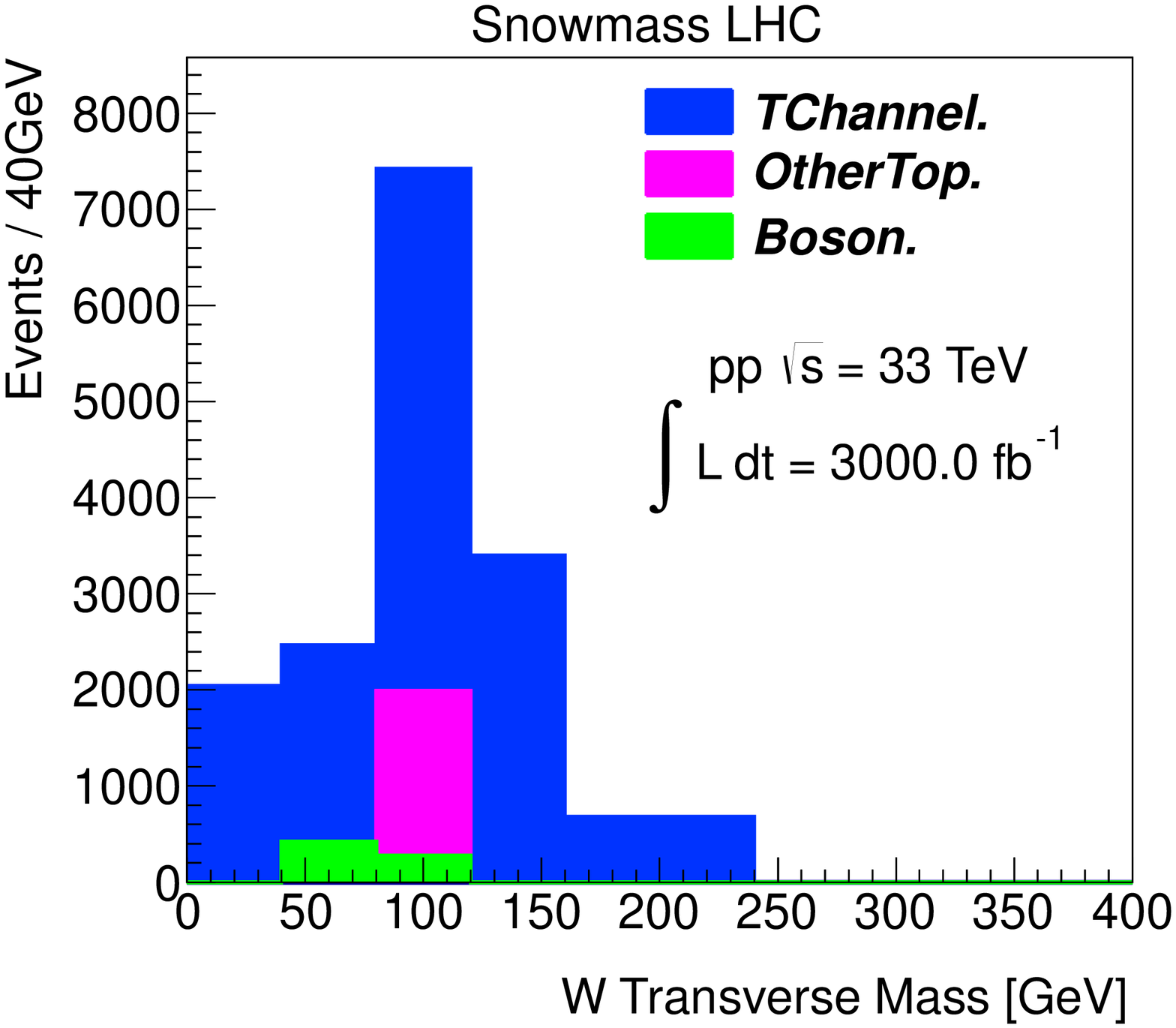}
    \label{fig:a}
  }
  \hspace*{0.0\textwidth}
  \subfigure[]{
    \includegraphics[width=0.40\textwidth]{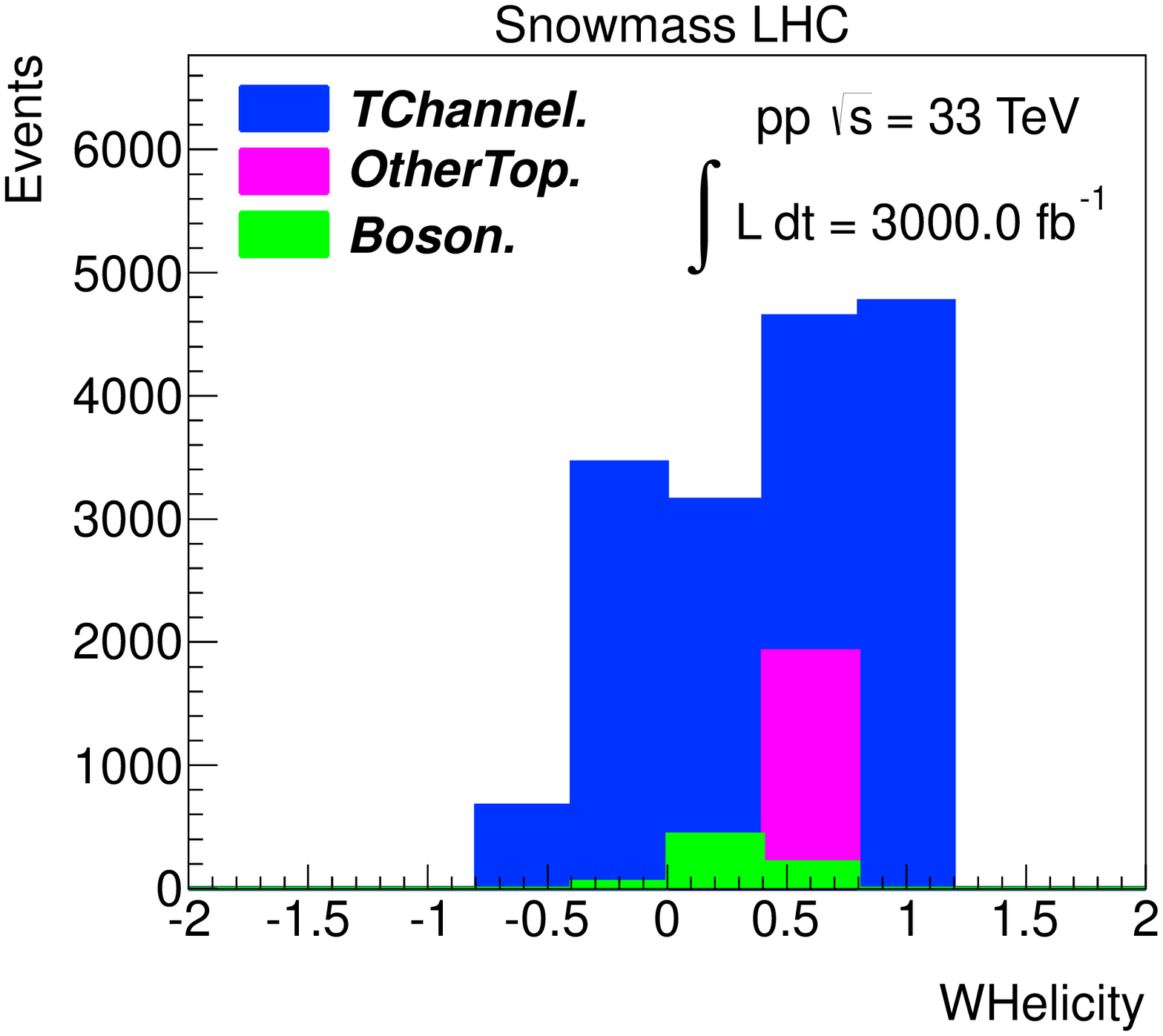}
    \label{fig:b}
  }
  \subfigure[]{
    \includegraphics[width=0.40\textwidth]{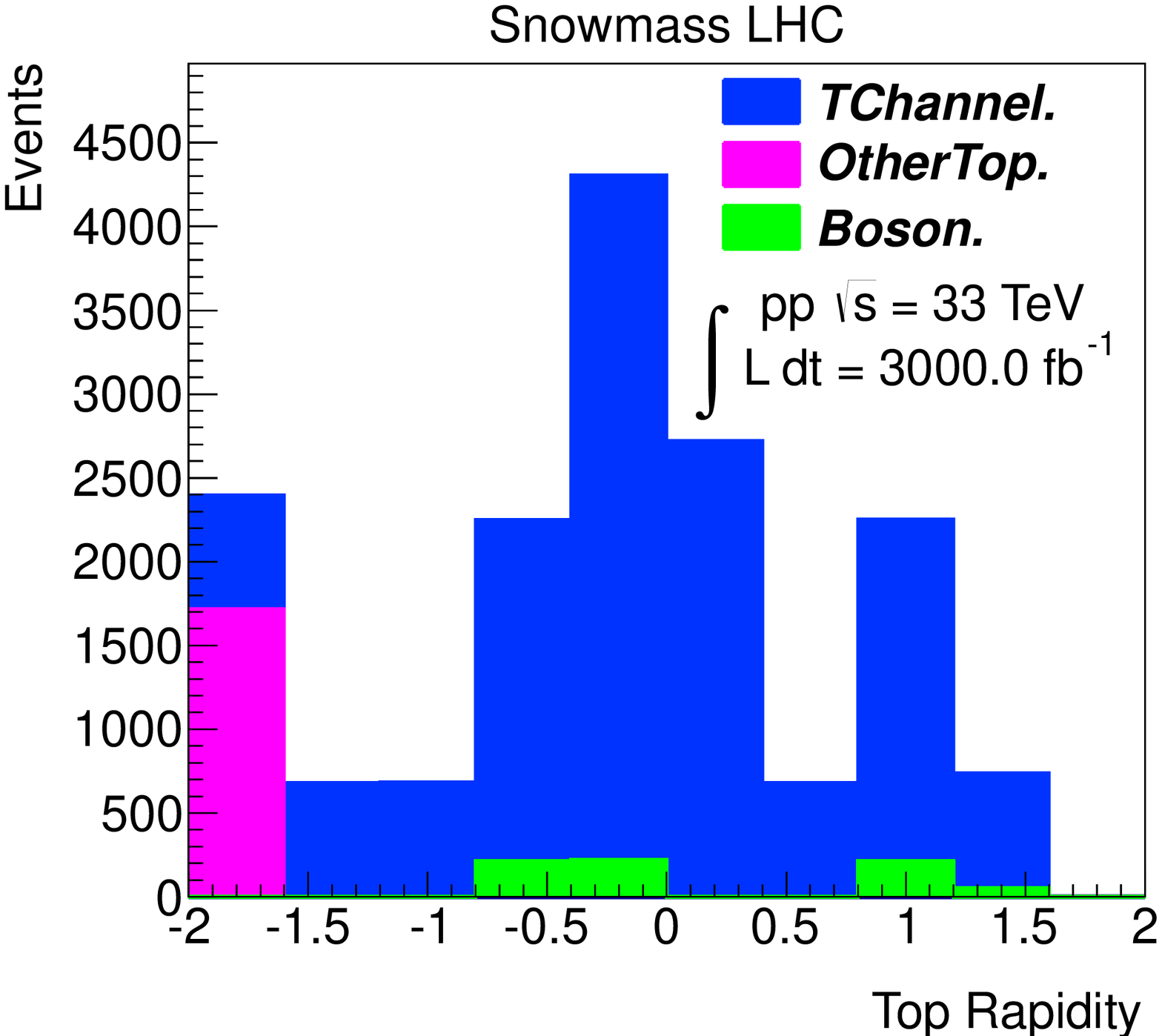}
    \label{fig:c}
  } 
  \hspace*{0.0\textwidth}
  \subfigure[]{
    \includegraphics[width=0.40\textwidth]{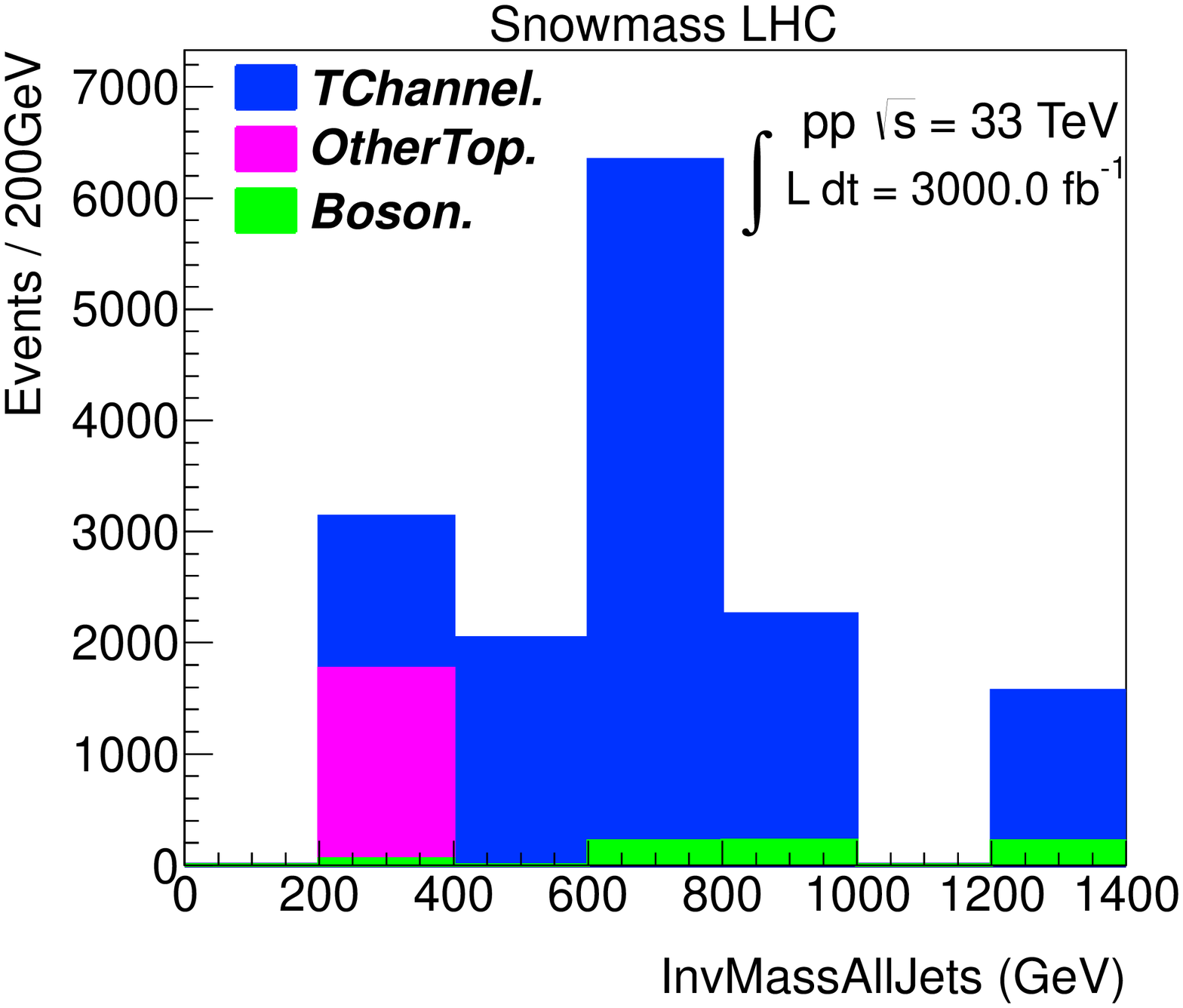}
    \label{fig:d}
  } 
  \caption{Kinematic distributions for events passing selection cuts for 3000~fb$^{-1}$
at 33~TeV: (a) transverse mass of the W boson, (b) helicity of the W boson, (c) rapidity of
the top quark  and (d) invariant mass of all the jets in the event.}
  \label{fig:afterkinematics33}
\end{figure}

Figures~\ref{fig:afterkinematics} and~\ref{fig:afterkinematics33} show kinematic distributions
after all cuts for 300~fb$^{-1}$ at 14~TeV and 3000~fb$^{-1}$ 33~TeV, respectively.

%
%
\section{Conclusions}
\label{sec:conclusions}
We have presented the $t$-channel single top quark production cross-section measurement
for three scenarios, studied within the context of the Snowmass energy frontier group:
300~fb$^{-1}$ of 14~TeV $pp$ data with an average pileup of 50~events,
3000~fb$^{-1}$ of 14~TeV $pp$ data with an average pileup of 140~events,
and 3000~fb$^{-1}$ of 33~TeV $pp$ data. For each the appropriate Snowmass detector models
are used. The $t$-channel events are selected in the lepton+jets top quark decay mode.
The background consists mainly of $W$+jets and top quark pair events. The $t$-channel signal
is isolated through a series of cuts, resulting in large, pure event samples.
The expected $t$-channel cross-section precision is about 5\% or better.

\newpage

\begin{acknowledgments}
We acknowledge the support of the entire Snowmass effort, in particular the samples
provided by the energy frontier. This work was supported in part by the U.S. National
Science Foundation under Grants
No. PHY-0952729 and PHY-1068318.
\end{acknowledgments}

\bibliography{tchan}

\end{document}